\documentclass[floatfix,twocolumn,showpacs,preprintnumbers,amsmath,amssymb,pra,superscriptaddress,longbibliography]{revtex4-1}
\usepackage{color}
\usepackage[usenames,dvipsnames,svgnames,table]{xcolor}
\usepackage[colorlinks=true,linkcolor=blue,urlcolor=blue,citecolor=blue]{hyperref}

\usepackage{mathtools}
\usepackage{graphicx}
\usepackage{dcolumn}
\usepackage{array}
\usepackage{lipsum}
\usepackage{bm}
\usepackage{subfigure}
\usepackage{amssymb}
\usepackage{multirow}
\usepackage{tabularx}
\usepackage{amsmath}
\usepackage{braket}
\graphicspath{{plots/}}

\begin{document}

\title[]{Finite-size effects in the reconstruction of dynamic properties from \textit{ab initio} path integral Monte-Carlo simulations}

\author{Tobias Dornheim}
\address{Center for Advanced Systems Understanding (CASUS), D-02826 G\"orlitz, Germany}
\email{t.dornheim@hzdr.de}

\author{Jan Vorberger}
\address{Helmholtz-Zentrum Dresden-Rossendorf (HZDR), 01328 Dresden, Germany}

\begin{abstract}
We systematically investigate finite-size effects in the dynamic structure factor $S(q,\omega)$ of the uniform electron gas obtained via the analytic continuation of \textit{ab initio} path integral Monte-Carlo (PIMC) data for the imaginary-time density--density correlation function $F(q,\tau)$. Using the recent scheme by Dornheim \textit{et al.}~[PRL \textbf{121}, 255001 (2018)], we find that the reconstructed spectra are not afflicted with any finite-size effects for as few as $N=14$ electrons both at warm dense matter (WDM) conditions and at the margins of the strongly correlated electron liquid regime. Our results further corroborate the high quality of our current description of the dynamic density response of correlated electrons, which is of high importance for many applications in WDM theory and beyond.
\end{abstract}

\maketitle

\section{Introduction\label{sec:introduction}}

Over the recent years, there has been a high interest in the study of matter at extreme conditions~\cite{fortov_review}. Of particular importance is so-called warm dense matter (WDM)~\cite{wdm_book}, a remarkable state with high densities ($r_s=a/a_\textnormal{B}\sim1$, with $a$ and $a_\textnormal{B}$ being the average interparticle distance and first Bohr radius) and temperatures ($\theta=k_\textnormal{B}T/E_\textnormal{F}\sim1$, with $E_\textnormal{F}$ being the usual Fermi energy). These conditions are well-known to occur in astrophysical objects like giant planet interiors~\cite{saumon1,Militzer_2008,Guillot2018} and brown dwarfs~\cite{becker,saumon1}. Moreover, they are expected to occur on the pathway towards inertial confinement fusion~\cite{hu_ICF}, which promises a potential abundance of clean energy in the future. Consequently, WDM research constitutes a topical frontier at the intersection of plasma physics and materials science, and WDM is nowadays routinely realized in large research centers around the globe such as NIF~\cite{Moses_NIF} and LCLS~\cite{LCLS_2016} in California, or the brandnew European X-FEL~\cite{Tschentscher_2017} in Germany. Other techniques include diamond anvil cells and heavy-ion beams, see Ref.~\cite{falk_wdm} for a recent review article on WDM experiments. Indeed, there have been many spectacular recent discoveries, e.g., Refs.~\cite{Kraus2016,Kraus2017,Kritcher69,Mo1451}.

Yet, the theoretical description of WDM is notoriously difficult due to the intriguingly intricate interplay of i) Coulomb coupling, ii) quantum degeneracy effects, and iii) thermal excitations~\cite{new_POP,wdm_book,review}. This renders WDM theory a marvellous challenge, and to this date there does not exist a single method that is capable to provide an accurate and reliable description of WDM applications in all situations. Among the most promising approaches are quantum Monte-Carlo (QMC) methods, such as the well-known path integral Monte-Carlo (PIMC) technique~\cite{cep,Takahashi,Berne_PIMC}. On the one hand, QMC methods can take into account the effects i)-iii) without any approximations and, thus, are capable to provide highly accurate input for other simulation methods. On the other hand, QMC simulations of electrons are severely limited by the notorious fermion sign problem (FSP), which leads to an exponential increase in computation time with decreasing temperature or increasing system size~\cite{troyer,dornheim_sign_problem,Dornheim_JCP_2019}. 
For these reasons, there has been a remarkable spark of new developments concerning the QMC simulation of fermions at finite temperature~\cite{Driver_PRL_2012,Brown_PRL,Blunt_PRB_2014,Schoof_PRL,Filinov_PRE_2015,Chin_PRE_2015,Dornheim_NJP_2015,Dornheim_JCP_2015,Militzer_PRL_2015,Malone_JCP_2015,Groth_PRB_2016,Dornheim_PRB_2016,Malone_PRL_2016,dornheim_prl,dornheim_cpp,groth_prl,Clark_PRB_2017,Dornheim_POP_2017,Liu2018,Dornheim_CPP_2019,Spencer2019,2020arXiv200310317H,Dornheim_PRL_2020,RCPIMC,dornheim2020attenuating}. 
 
An important mile stone was given by the construction of the first accurate QMC-based parametrizations of the exchange--correlation free energy $f_\textnormal{xc}$ of the uniform electron gas (UEG)~\cite{Karasiev_PRL_2014,groth_prl}, which allow for the possibility to perform density functional theory (DFT) calculations of WDM on the level of the local density approximation. Indeed, it has been revealed in several independent studies by different groups that using $f_\textnormal{xc}$---in contrast to the usual \emph{ground-state approximation} where the temperature-dependent $f_\textnormal{xc}$ is replaced by the zero-temperature limit $\lim_{T\to0}f_\textnormal{xc}=\epsilon_\textnormal{xc}$~\cite{Ceperley_Alder_1980,Vosko_Wilk_Nusair,Perdew_Zunger}---that temperature-effects in the exchange--correlation functional cannot be neglected in the WDM regime~\cite{Sjostrom_PRB_2014,karasiev_importance,kushal}.

Another important field of investigation is the electronic density response to an external perturbation. Within linear response theory~\cite{quantum_theory}, this is fully characterized by the dynamic density response function~\cite{kugler1}
\begin{eqnarray}\label{eq:chi}
\chi(q,\omega) = \frac{\chi_0(q,\omega)}{1-\tilde v(q)\left(1-G(q,\omega)\right)\chi_0(q,\omega) } \quad .
\end{eqnarray}
Here, $\chi_0(q,\omega)$ denotes the density response function of the ideal Fermi gas and the dynamic local field correction (LFC) $G(q,\omega)$ contains the full wave-number- and frequency-resolved description of exchange--correlation effects in the system. For example, setting $G(q,\omega)=0$ in Eq.~(\ref{eq:chi}) corresponds to a mean-field description of the dynamic density response, and is typically referred to as the \emph{random phase approximation} (RPA). Naturally, the information contained in $G(q,\omega)$ is vital for many applications, like the construction of advanced exchange-correlation functionals for DFT~\cite{Lu_JCP_2014,Thygesen_JCP_2015,Burke_PRL_2016,Goerling_PRB_2019} and time-dependent DFT~\cite{Baczewski_PRL_2016}, including electronic correlations into quantum hydrodynamics~\cite{Moldabekov_POP_2018,new_POP,Diaw_PRE_2015,Diaw2017},
taking into account electronic screening into effective ion--ion potentials~\cite{Ceperley_potential,zhandos1,zhandos2}, and the computation of many physical observables like electrical and thermal conductivities~\cite{Veysman_PRE_2016,Desjarlais_PRE_2017}, energy loss characteristics~\cite{Moldabekov_PRE_2020}, and energy transfer rates~\cite{transfer1,transfer2}.
Moreover, we mention the interpretation of X-ray Thomson scattering (XRTS) experiments~\cite{siegfried_review,dornheim2020effective}, e.g. within the framework of the Chihara decomposition~\cite{Chihara_1987,kraus_xrts}---a de-facto standard method of diagnostics in WDM experiments.

Unfortunately, QMC methods are inherently incapable to directly compute time-dependent (or, equivalently, frequency-dependent) properties due to an additional dynamical sign problem~\cite{Dynamical_Sign_Problem,Dynamical_Sign_Problem2}. Therefore, the first accurate results for the electronic density response have been obtained in the static limit (i.e., $\omega\to0$) based on the ground-state QMC simulations by Moroni \textit{et al.}~\cite{moroni,moroni2,cdop}. Very recently, Dornheim \textit{et al.}~\cite{dornheim_ML} were able to extend this description to finite temperature by presenting extensive new PIMC results for the static LFC over a broad parameter range. These new data were subsequently used to train a fully connected deep neural network, which provides an accurate description of the static LFC $G(q):=G(q,\omega=0)$ (and, thus, also $\chi(q)$ etc.) covering the entire WDM regime ($0.7\leq r_s\leq 20$ and $0\leq\theta\leq4$). Furthermore, the same group also presented similar data for the static density response both for the strongly coupled electron liquid~\cite{dornheim_electron_liquid} ($20\leq r_s \leq 100$) and the weakly coupled high energy density limit regime~\cite{dornheim_HEDP} ($0.05\leq r_s \leq0.5$). Lastly, we mention that accurate PIMC data for the static response have become available even for the nonlinear regime~\cite{Dornheim_PRL_2020}.

On the other hand, an unbiased \textit{ab initio} description of the full frequency dependence of either $\chi(q,\omega)$ or, equivalently, $G(q,\omega)$ is most challenging. For example, the nonequilibrium Green function method~\cite{kwong_prl-00,Kas_PRL_2017} is based on a perturbative expansion around the noninteracting system and, thus, cannot fully take into account the effects due to electronic correlations that are important in the WDM regime. Other approximate methods include the extension of a dynamical mean-field description by using known static limits for the exchange--correlation effects within the frame-work of the method of frequency moments by Tkachenko and co-workers~\cite{Tkachenko_PRL_2012,Tkachenko_PRL_2017,Tkachenko_CPP_2018}, or the ground-state many-body approach by Takada~\cite{Takada_PRB_2016,Takada_PRL_2002}. Yet, the accuracy of these methods had remained unclear, and reliable benchmark data were highly needed.

While real-time dependent simulations still remain out of reach, there does exist a neat alternative: the analytic continuation of an imaginary-time correlation function. More specifically, the PIMC method allows to obtain exact results for the imaginary-time density--density correlation function $F(q,\tau)$ [cf.~Eq.~(\ref{eq:define_F}) below], which can be used as input for the reconstruction of dynamic properties. The required inverse Laplace transform is a well-known, but notoriously difficult problem~\cite{Jarrell_PhysRep_1996,Goulko_Analytic_Continuation_PRB_2017} as the reconstructed spectra might not be unique, and the problem statement is ill-posed with respect to the inevitable Monte-Carlo error bars. This problem was recently solved for the specific case of the dynamic density response of the UEG by Dornheim \textit{et al.}~\cite{dornheim_dynamic,dynamic_folgepaper}, who were able to obtain the first PIMC data for the dynamic structure factor $S(q,\omega)$ going from WDM conditions ($r_s=2$) to the margins of the electron liquid regime ($r_s=10$). These new results have opened up many avenues for future investigations, like the investigation and possible experimental verification of an incipient excitionic mode that appears with increasing electronic correlation effects~\cite{dornheim_dynamic,Takada_PRB_2016}. Another hands-on application of the results from Refs.~\cite{dornheim_dynamic,dynamic_folgepaper} would be the construction of a dynamic exchange--correlation kernel for time-dependent DFT simulations~\cite{Baczewski_PRL_2016}. 
Yet, one detail of this new approach to the dynamic properties of WDM has remained unaddressed: the finite size of the simulation cell in any PIMC simulation.

Typically, these calculations use $N=30\dots100$ electrons, and the respective PIMC data explicitly depend on the system size. For example, to obtain accurate results for the interaction energy per particle $v$ in the thermodynamic limit (i.e., $\lim_{N\to\infty}V/N$), even $N\sim10^2-10^3$ electrons are not necessarily sufficient~~\cite{dornheim_prl,Dornheim_POP_2017,review}. In contrast, wave-number resolved quantities like the static structure factor $S(q)$ or the static density response function $\chi(q)$ are known empirically to converge much faster with $N$~\cite{dornheim_pre,groth_jcp,dornheim_prl,dornheim_ML}. Moreover, finite-size effects in $\chi(q)$ can even be removed from the QMC data by applying a subsequent finite-size correction~\cite{groth_jcp,moroni,moroni2,dornheim_ML,dornheim_electron_liquid,dornheim_HEDP}.

In this work, we verify that these findings do indeed also hold for the reconstructed results for $S(q,\omega)$ from Refs.~\cite{dornheim_dynamic,dynamic_folgepaper}. To this end, we have carried out extensive \textit{ab initio} PIMC simulations of the UEG for different system sizes $N=8,\dots,100$ for two relevant parameter combinations: a) WDM conditions with $r_s=2$ (metallic density) and $\theta=2$, and b) the margins of the electron liquid regime with $r_s=10$ and $\theta=1$. 
Conditions as in case a) are more relevant as they are close to states of matter as in state-of-the-art experiments. Yet, the impact of dynamic local field effects is limited and exhausts itself in a red-shift of the dispersion relation compared to RPA. However, the UEG at conditions b) exhibits a highly interesting behaviour, with a negative dispersion relation and non-trivial double-peak structures in $S(q,\omega)$ at intermediate wave numbers. Moreover, it has been shown in Ref.~\cite{dornheim_prl} that the full frequency-dependence of $G(q,\omega)$ must be taken into account to get accurate results for $S(q,\omega)$. 
Therefore, our current investigation of both regimes a) and b) will be very useful both for practical WDM applications like the construction of XC-kernels for TD-DFT, and for theoretical challenges like understanding the physics behind a possible new mode at large $r_s$.

This paper is organized as follows: in Sec.~\ref{sec:theory}, we introduce the required theoretical background covering both the path integral Monte-Carlo method (Sec.~\ref{sec:pimc}) and the reconstruction of the dynamic structure factor (Sec.~\ref{sec:reconstruction}). We start our investigation in Sec.~\ref{sec:results} with a brief discussion of the fermion sign problem and then investigate finite-size effects both in the PIMC input data and the reconstructed dynamic structure factors both at WDM conditions (Sec.~\ref{sec:WDM}) and at the margins of the strongly correlated electron liquid regime (\ref{sec:EL}). The paper is concluded by a brief summary and outlook in Sec.~\ref{sec:summary}.

Note that we assume Hartree atomic units throughout this work.

\section{Theory\label{sec:theory}}

\subsection{Path integral Monte-Carlo\label{sec:pimc}}

The path integral Monte-Carlo method (see Ref.~\cite{cep} for an extensive review article) is based on a stochastic evaluation of the canonical (i.e., particle number $N$, temperature $T$, and volume $V=L^3$ are fixed) density matrix evaluated in coordinate space,
\begin{eqnarray}\label{eq:Z}
 Z = \frac{1}{N_\uparrow!N_\downarrow!} \sum_{\sigma_\uparrow\in S_{N_{\uparrow}}}\sum_{\sigma_\downarrow\in S_{N_{\downarrow}}} \textnormal{sgn}^\textnormal{f}(\sigma_\uparrow,\sigma_\downarrow) 
 \\ \nonumber
 \int \textnormal{d}\mathbf{R}\ \bra{ \mathbf{R} } e^{-\beta\hat{H}} \ket{ \hat{\pi}_{\sigma_\uparrow}\hat{\pi}_{\sigma_\downarrow}\mathbf{R}} \quad .
\end{eqnarray}
Here $N_\uparrow=N_\downarrow=N/2$ denote the number of spin-up and -down electrons, and the double sum over the respective permutation groups $S_i$ are needed for a proper antisymmetrization, i.e., to take into account Fermi statistics~\cite{dornheim_sign_problem,Dornheim_JCP_2019}. Moreover, $\hat\pi_i$ are the permutation operators for a particular element from each group, and the sign function $\textnormal{sgn}^\textnormal{f}(\sigma_\uparrow,\sigma_\downarrow)$ is positive (negative) for an even (odd) number of pair permutations. At this point, we note that a complete introduction and derivation to the PIMC method has already been presented elsewhere~\cite{cep,Berne_PIMC,Takahashi} and does not have to be repeated here.

For the present purpose, it is fully sufficient to work with the abstract expression
\begin{eqnarray}\label{eq:weights}
Z = \int \textnormal{d}\mathbf{X}\ W(\mathbf{X})\ ,
\end{eqnarray}
which can be interpreted in the following way: the full partition function has been recast into an integration over all possible paths $\mathbf{X}$ in the imaginary time (see, e.g., Ref.~\cite{dornheim_sign_problem} for examples and graphical depictions), and each path has to be taken into account with the appropriate weight $W(\mathbf{X})$. For bosons and boltzmannons (i.e., distiguishable particles~\cite{Dornheim_CPP_2016,Clark_Casula_Ceperley_Hexatic}), $W(\mathbf{X})$ is strictly positive and it is straightforward to use the Metropolis algorithm~\cite{Metropolis} to generate a Markov chain of random configurations that are distributed according to the probability $P(\mathbf{X})=W(\mathbf{X})/Z$. For fermions, on the other hand, $W(\mathbf{X})$ can be both positive and negative [cf.~Eq.~(\ref{eq:Z})], and $P(\mathbf{X})$ cannot be interpreted as a proper probability distribution.

As a practical workaround, we instead generate the paths according to the modulus value of the weights, $W'(\mathbf{X})=|W(\mathbf{X})|$. It is easy to see that the fermionic expectation value of an arbitrary observable $\hat A$ is then given by
\begin{eqnarray}\label{eq:ratio}
\braket{\hat A} = \frac{\braket{\hat A\hat S}'}{\braket{\hat S}'}\ ,
\end{eqnarray}
where $\braket{\dots}'$ denotes the expectation value taken with respect to the modified weights $W'$, and $S(\mathbf{X})=W(\mathbf{X})/|W(\mathbf{X})|$ is the estimator for the sign. 
The denominator of Eq.(\ref{eq:ratio}) is commonly known as the \emph{average sign} $S$ and constitutes a measure for the amount of cancellations due to positive and negative terms within a fermionic PIMC simulation.
More specifically, the sign scales as
\begin{eqnarray}\label{eq:exp}
S = \textnormal{exp}\left(-\beta N (f-f')
\right)\ ,
\end{eqnarray}
where $f$ and $f'$ denote the free energy densities of the original and the modified systems, respectively.
This is highly problematic as the statistical uncertainty $\Delta A$ of a fermionic PIMC expectation value [Eq.~(\ref{eq:ratio})] is inversely proportional to $S$, and, thus, exponentially increases with increasing the system size $N$ or increasing the inverse temperature $\beta=1/k_\textnormal{B}T$,
\begin{eqnarray}\label{eq:error}
\frac{\Delta A}{A} \sim \frac{1}{S\sqrt{N_\textnormal{MC}}} \sim \frac{\textnormal{exp}\left(-\beta N (f-f')
\right)}{\sqrt{N_\textnormal{MC}}}\ .
\end{eqnarray}
Evidently, the error bar can only be decreased by increasing the number of Monte-Carlo samples as $1/\sqrt{N_\textnormal{MC}}$, which quickly becomes unfeasible. Therefore, Eq.~(\ref{eq:error}) constitutes an \emph{exponential wall} with respect to $N$ and $\beta$ that is being referred to as the fermion sign problem~\cite{dornheim_sign_problem,Loh_sign_problem,Lyubartsev_2005,troyer}.

For completeness, we mention that all PIMC simulations in this work have been carried out using a canonical adaption~\cite{mezza} of the worm algorithm introduced by Boninsegni \textit{et al.}~\cite{boninsegni1,boninsegni2}.

\subsection{Reconstruction of dynamic properties\label{sec:reconstruction}}

As a side effect of its formulation in imaginary time, the PIMC method allows for a straightforward evaluation of a variety of imaginary-time correlation functions, such as the Matsubara Green function~\cite{boninsegni1,Filinov_PRA_2012} 
or the velocity autocorrelation function~\cite{Rabani1129}. In this work, we are interested in the imaginary-time version of the intermediate scattering function
\begin{eqnarray}\label{eq:define_F}
F(q,\tau) = \frac{1}{N}  \braket{\hat\rho(q,\tau)\hat\rho(-q,0)}\,,
\end{eqnarray}
which is nothing else than the density--density correlation function evaluated at an imaginary-time argument $\tau\in[0,\beta]$, see, e.g., Refs.~\cite{dynamic_folgepaper,dornheim_ML,Filinov_PRA_2012,Boninsegni1996,Motta_JCP_2015} for a few examples.

One practical application of Eq.~(\ref{eq:define_F}) is its relation to the static density response function, which is simply given by a one-dimensional integral over the $\tau$-axis~\cite{Bowen_PRB_1992},
    \begin{eqnarray}\label{eq:static_chi}
\chi({q},0) = -n\int_0^\beta \textnormal{d}\tau\ F({q},\tau)\ .
\end{eqnarray}
In fact, this relation was paramount for our current understanding of the static density response of correlated electrons at finite temperature~\cite{dornheim_ML,dornheim_electron_liquid,dornheim_HEDP} as it allows to obtain the complete wave-number description of $\chi$ from a single simulation of the unperturbed system.

In the present work, we focus on the relation
\begin{eqnarray}\label{eq:S_F}
F(q,\tau) = \int_{-\infty}^\infty \textnormal{d}\omega\ S(q,\omega) e^{-\tau\omega} \quad ,
\end{eqnarray}
which means that $F(q,\tau)$ is connected to the dynamic structure factor via a Laplace transform. The problem statement is thus to solve Eq.~(\ref{eq:S_F}) for $S(q,\omega)$ by numerically performing an inverse Laplace transform, which is a notoriously hard and, in fact, ill-posed problem~\cite{Jarrell_PhysRep_1996}.
The main obstacle is given by the fact that
the PIMC data for Eq.~(\ref{eq:define_F}) are afflicted with a statistical error [cf.~Eq.~(\ref{eq:error})]. Therefore, there could potentially exist an infinite number of possible trial solutions $S_\textnormal{trial}(q,\omega)$ that, when being inserted into Eq.~(\ref{eq:S_F}), reproduce the PIMC values for $F(q,\tau)$ for all $\tau$-points within the given confidence interval.
To somewhat constrain the space of possible trial solutions, one can make use of the frequency moments of $S(q,\omega)$,
\begin{eqnarray}\label{eq:moments}
\braket{\omega^k} = \int_{-\infty}^\infty \textnormal{d}\omega\ S(q,\omega) \omega^k \ ,
\end{eqnarray}
with the cases $k=-1,0,1,3$ being known from different sum-rules, see Ref.~\cite{dynamic_folgepaper} for a detailed overview.

Over the years, many reconstruction methods have been proposed, including genetic algorithms~\cite{Vitali_PRB_2010,Bertaina_GIFT_2017}, maximum entropy methods~\cite{Jarrell_PhysRep_1996,Boninsegni_maximum_entropy,Fuchs_PRE_2010}, Monte-Carlo sampling~\cite{Mishchenko_PRB_2000,Filinov_PRA_2012}, or machine-learning schemes~\cite{ML_analytic_continuation}, see Ref.~\cite{Schoett_PRB_2016} for a recent comparison of different methods.
More specifically, the reconstruction of the dynamic structure factor starting from Eq.~(\ref{eq:S_F}) has allowed for profound insights into the physics of, e.g., ultracold atoms like $^4$He~\cite{Boninsegni1996,Vitali_PRB_2010} or quantum-dipole systems~\cite{Filinov_PRA_2012} and even supersolids~\cite{Saccani_Supersolid_PRL_2012}. Yet, for the case of the warm dense electron gas, the combined information within $F(q,\tau)$ and $\braket{\omega^k}$ did still not sufficiently constrain the space of possible trial solutions $S_\textnormal{trial}(q,\omega)$, and additional input was needed.

To overcome this obstacle, Dornheim \textit{et al.}~\cite{dornheim_dynamic} proposed to invoke the fluctuation--dissipation theorem~\cite{quantum_theory}
\begin{eqnarray}\label{eq:FDT}
S(\mathbf{q},\omega) = - \frac{ \textnormal{Im}\chi(\mathbf{q},\omega)  }{ \pi n (1-e^{-\beta\omega})}\ ,
\end{eqnarray}
which states that the dynamic structure factor is fully defined by the dynamic density response function introduced in Eq.~(\ref{eq:chi}). Moreover, we have already mentioned that the only unknown part of $\chi(q,\omega)$ is the dynamic local field correction $G(q,\omega)$. In this way, the reconstruction of $S(q,\omega)$ has been recast into the quest for $G(q,\omega)$.

This has turned out highly advantageous, because many additional exact properties of the dynamic LFC are known in advance~\cite{dynamic_folgepaper}:
\begin{enumerate}
    \item The Kramers-Kronig relations between Re$G(q,\omega)$ and Im$G(q,\omega)$ allow to compute one from the other in both directions~\cite{kugler1}.
    \item It is known that Re$G(q,\omega)$ [Im$G(q,\omega)$] is an even [odd] function with respect to the frequency $\omega$.
    \item It holds Im$G(q,0)=\textnormal{Im}G(q,\infty)=0$.
    \item The exact static limit of Re$G(q,\omega)$ can be easily obtained from Eq.~(\ref{eq:static_chi}) and is also available as a convenient neural-net representation~\cite{dornheim_ML}.
    \item The high-frequency asymptotic of Re$G(q,\omega)$ is given by
    \begin{eqnarray}\label{eq:Ginfty}
\textnormal{Re}G(q,\infty) = I(q) - \frac{2q^2 K_\textnormal{xc}}{\omega^2_\textnormal{pl}}\ ,
\end{eqnarray}
where the exchange--correlation contribution to the kinetic energy $K_\textnormal{xc}$ is obtained from the parametrization by Groth \textit{et al.}~\cite{groth_prl}, and the interaction integral is defined as
 \begin{eqnarray}\label{eq:I}
 I(q) &=&  \frac{1}{8\pi^2n} \int_0^\infty \textnormal{d}k\ k^2 \big(1-S(k)\big) \\ \nonumber
 &\times& \Bigg( 
 \frac{5}{3} - \frac{k^2}{q^2} + \frac{\big(k^2-q^2\big)^2}{2kq^3} \textnormal{log}\Bigg| \frac{k+q}{k-q} \Bigg| 
 \Bigg) \quad .
 \end{eqnarray}

\end{enumerate}

The new reconstruction procedure from Refs.~\cite{dornheim_dynamic,dynamic_folgepaper} is based on a stochastic sampling of trial solutions $G_\textnormal{trial}(q,\omega)$, with the above exact properties being automatically satisfied. These are subsequently substituted into Eq.~(\ref{eq:chi}) to obtain the corresponding $\chi_\textnormal{trial}(q,\omega)$, and the fluctuation--dissipation theorem [Eq.~(\ref{eq:FDT})] allows to finally compute the trial solution for the dynamic structure factor, $S_\textnormal{trial}(q,\omega)$.

In the end, the $S_\textnormal{trial}(q,\omega)$ are then substituted into Eqs.~(\ref{eq:S_F}) and (\ref{eq:moments}) and compared to the PIMC data for both $F(q,\tau)$ and the sum-rule results for $\braket{\omega^k}$; those $S_\textnormal{trial}(q,\omega)$ that do not agree to these data within the given confidence interval are discarded.
The final solution for, e.g., the dynamic structure factor is then computed as the average over all $N_\textnormal{t}$ valid trial solutions $S_i(q,\omega)$
\begin{eqnarray}
S(q,\omega) = \frac{1}{N_\textnormal{t}}\sum_{i=1}^{N_\textnormal{t}} S_i(q,\omega)\ .
\end{eqnarray}
Moreover, this procedure allows for a straightforward estimation of the corresponding uncertainty as the variance of the $\{S_i(q,\omega)\}$
\begin{eqnarray} \label{eq:S_error}
\Delta S(q,\omega) = \left(
\frac{1}{N_\textnormal{t}} \sum_{i=1}^{N_\textnormal{t}} \left[
S(q,\omega) - S_i(q,\omega)
\right]^2
\right)^{1/2}\ .
\end{eqnarray}

\section{Results\label{sec:results}}

\begin{figure}\centering
\includegraphics[width=0.5\textwidth]{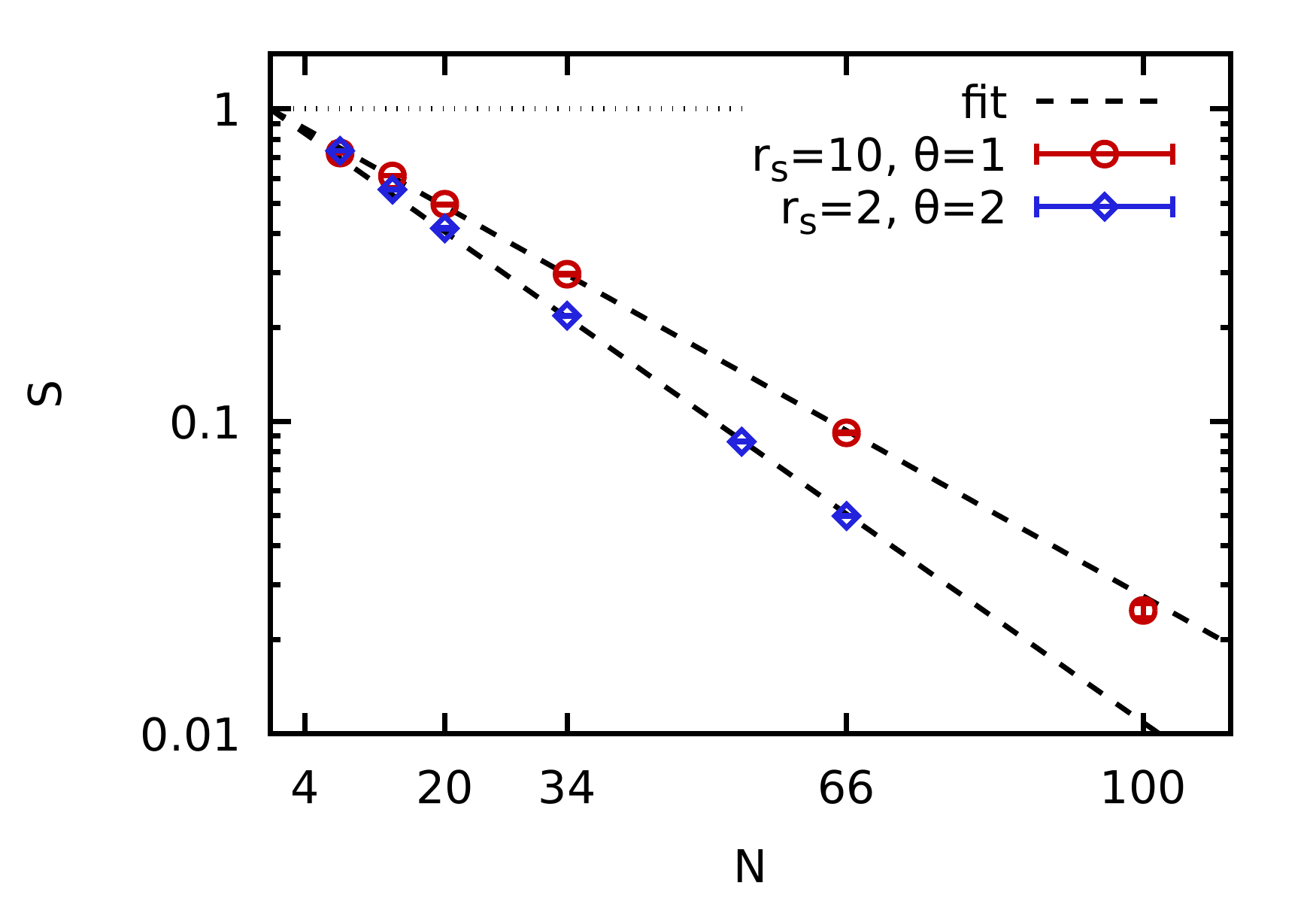}
\caption{\label{fig:Sign}
PIMC results for the average sign $S$ as a function of the system size $N$. The red circles and blue diamonds correspond to $r_s=10$, $\theta=1$ and $r_s=2$, $\theta=2$, respectively. The black dashed lines correspond to exponential fits according to Eq.~(\ref{eq:sign_fit}) for $N\geq20$. PIMC simulations are computationally feasible for $S\gtrsim 0.01$.
}
\end{figure}

Let us start our investigation of the finite-size effects in the reconstructed PIMC results for the dynamic structure factor $S(q,\omega)$ by briefly touching upon the fermion sign problem within the PIMC simulations for both parameter combinations to be discussed in this work. 
To this end, we show our PIMC data for the average sign $S$ [i.e., the denominator of Eq.~(\ref{eq:ratio})] as a function of the system size $N$ in Fig.~\ref{fig:Sign}. Here the red circles and blue diamonds correspond to electron liquid and WDM conditions, respectively and exhibit a qualitatively similar behaviour. More specifically, the sign is strictly monotonically decreasing with $N$, and the dashed black lines depict exponential fits according to
\begin{eqnarray}\label{eq:sign_fit}
S(N) = e^{-a N}\ ,
\end{eqnarray}
with $a$ being the free parameter. This is motivated by Eq.~(\ref{eq:exp}), and fits well to our PIMC data points for both cases. Furthermore, we note that the sign decays faster for the WDM case, which actually is a nontrivial finding. For higher densities (smaller values of $r_s$), electronic correlation effects become less important, and the particles are not that strongly separated. Consequently, fermionic exchange effects are more important, which results in a stronger amount of cancellations of positive and negative contributions, i.e., a decreasing average sign $S$. In this sense, the faster decrease of $S$ at $r_s=2$ surely is expected.

On the other hand, the electron liquid example has been obtained at half the value of the reduced temperature $\theta$. Let us for a moment consider the case of an ideal Fermi gas. In that case, the degree of quantum degeneracy, and, thus, the value of the corresponding average sign $S$ is fully defined by $\theta$ and does not depend on the density parameter $r_s$. Therefore, the sign for $r_s=10$ and $\theta=1$ would have been substantially lower compared to $r_s=2$ and $\theta=2$.
Hence, the slower decay of $S$ in the red circles for the interacting case is purely a result of the increased Coulomb repulsion upon decreasing the density.

For completeness, we note that data with a sufficient accuracy for the reconstruction of $S(q,\omega)$ can be obtained for $S\gtrsim0.1$ [resulting in a hundred-fold increase in computation time compared to bosons or boltzmannons, cf.~Eq.~(\ref{eq:error})], i.e., $N\lesssim54$ and $N\lesssim66$ electrons for the WDM and electron liquid examples.

\subsection{Warm dense matter regime: $r_s=2$ and $\theta=2$\label{sec:WDM}}

\begin{figure}\centering
\includegraphics[width=0.55\textwidth]{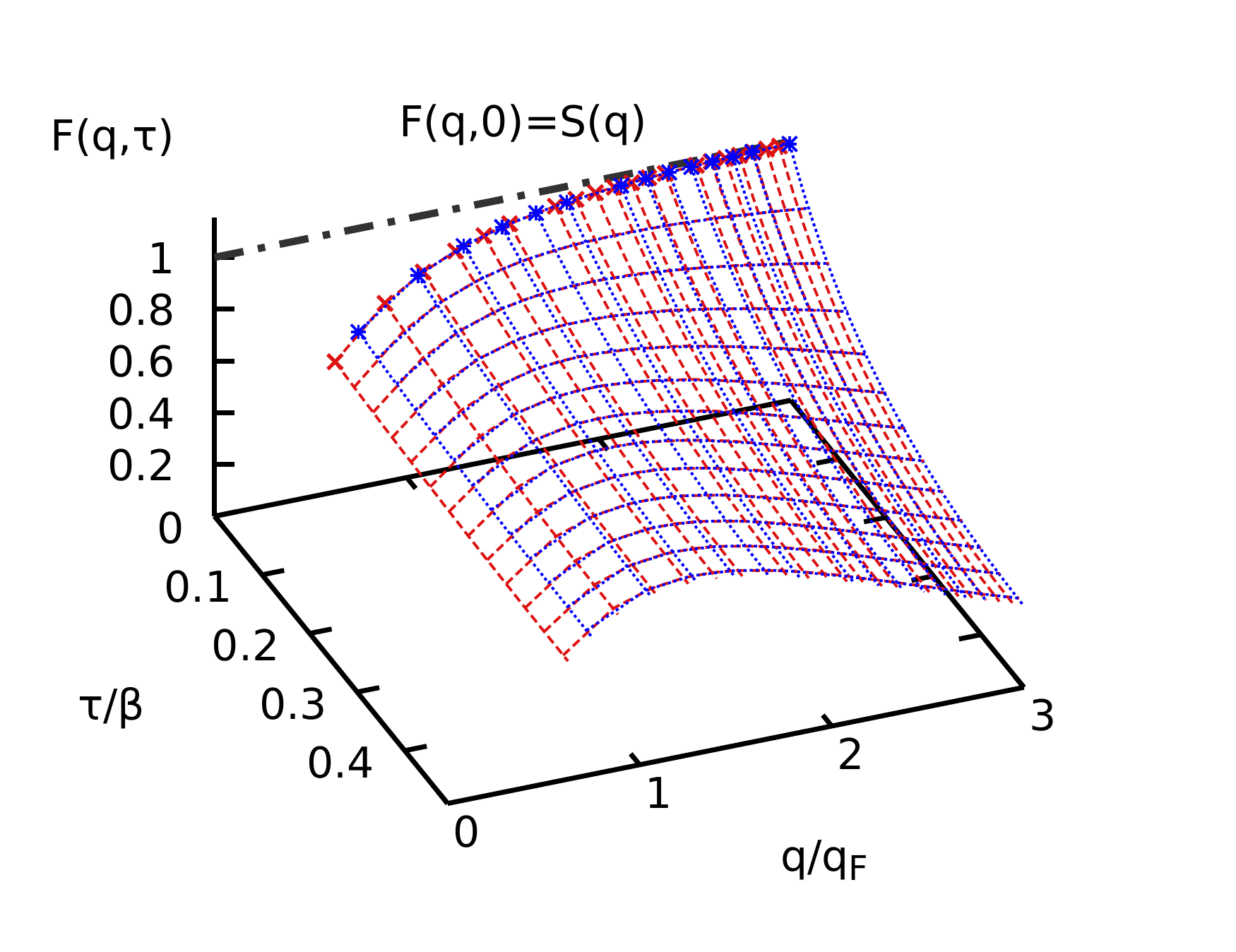}\vspace*{-0.7cm}
\caption{\label{fig:FPLOT_rs2_theta2}
PIMC data for the imaginary-time density--density correlation function $F$ [cf.~Eq.~(\ref{eq:define_F})] in the $\tau$-$x$-plane for $r_s=2$ and $\theta=2$. The red (dashed) and blue (dotted) data have been obtained for $N=34$ and $N=20$ electrons. Note that $F$ approaches the static structure factor in the $\tau=0$ limit (crosses and stars) and that it is symmetric with respect to $\tau=\beta/2$, i.e., $F(q,\tau)=F(q,\beta-\tau)$.
}
\end{figure}

Due to the high current interest in the WDM regime, we will start our investigation of the system-size dependence of the dynamic structure factor at $r_s=2$ and $\theta=2$. 
However, before we consider $S(q,\omega)$ itself, it makes sense to first investigate any finite-size effects in the quantities that are used as input for the reconstruction. The imaginary-time density--density correlation function constitutes the most important ingredient and is depicted in Fig.~\ref{fig:FPLOT_rs2_theta2} for $N=20$ (blue dotted) and $N=34$ (red dashed) unpolarized electrons in the $\tau$-$q$-plane. First and foremost, we note that a direct comparison between the two data sets is not possible, as $F$ is available at different $q$-points. This is an immediate consequence of the momentum quantization due to the finite box length $L$, see Refs.~\cite{dornheim_prl,dornheim_cpp} for a detailed explanation. The discretization in the $\tau$-direction, on the other hand, is defined by the selected number of imaginary-time propagators within a PIMC simulation, and can be chosen arbitrarily fine. In this work, we always use $P=200$ primitive propagators (see Refs.~\cite{Brualla_JCP_2004,Sakkos_JCP_2009,Dornheim_CPP_2019} for a detailed and accessible discussion), which is sufficient to ensure convergence and allows for an adequate resolution with respect to the imaginary time $\tau$. We note that not all $\tau$-points are depicted in Fig.~\ref{fig:FPLOT_rs2_theta2} to make it more accessible.

The red crosses and blue stars depict the $\tau\to0$ limit of $F(q,\tau)$, which is given by the usual static structure factor
\begin{eqnarray}\label{eq:Sq}
F(q,0) = S(q) = \int_{-\infty}^\infty \textnormal{d}\omega\ S(q,\omega)\ .
\end{eqnarray}
In other words, Eq.~(\ref{eq:Sq}) means that the normalization of the reconstructed $S(q,\omega)$ (or, equivalently, $\braket{\omega^0}$) is known in advance.

Let us next come to the topic at hand, which is the dependence of $F(q,\tau)$ on the number of electrons $N$. Although no difference between the two particle numbers can be seen in Fig.~\ref{fig:FPLOT_rs2_theta2}, the depicted surface plot is not optimal for this purpose.\begin{figure*}\centering
\includegraphics[width=0.475\textwidth]{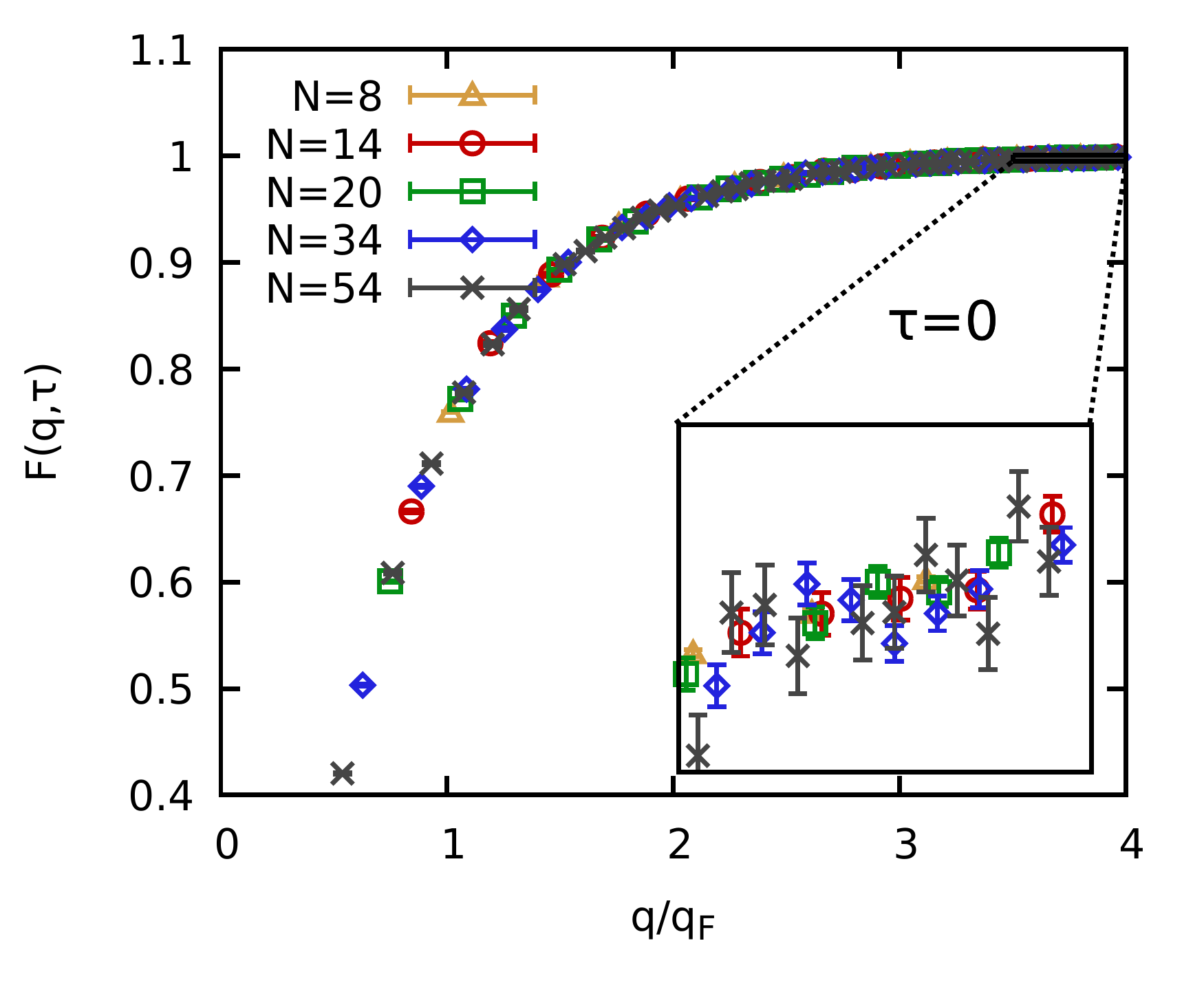}
\includegraphics[width=0.475\textwidth]{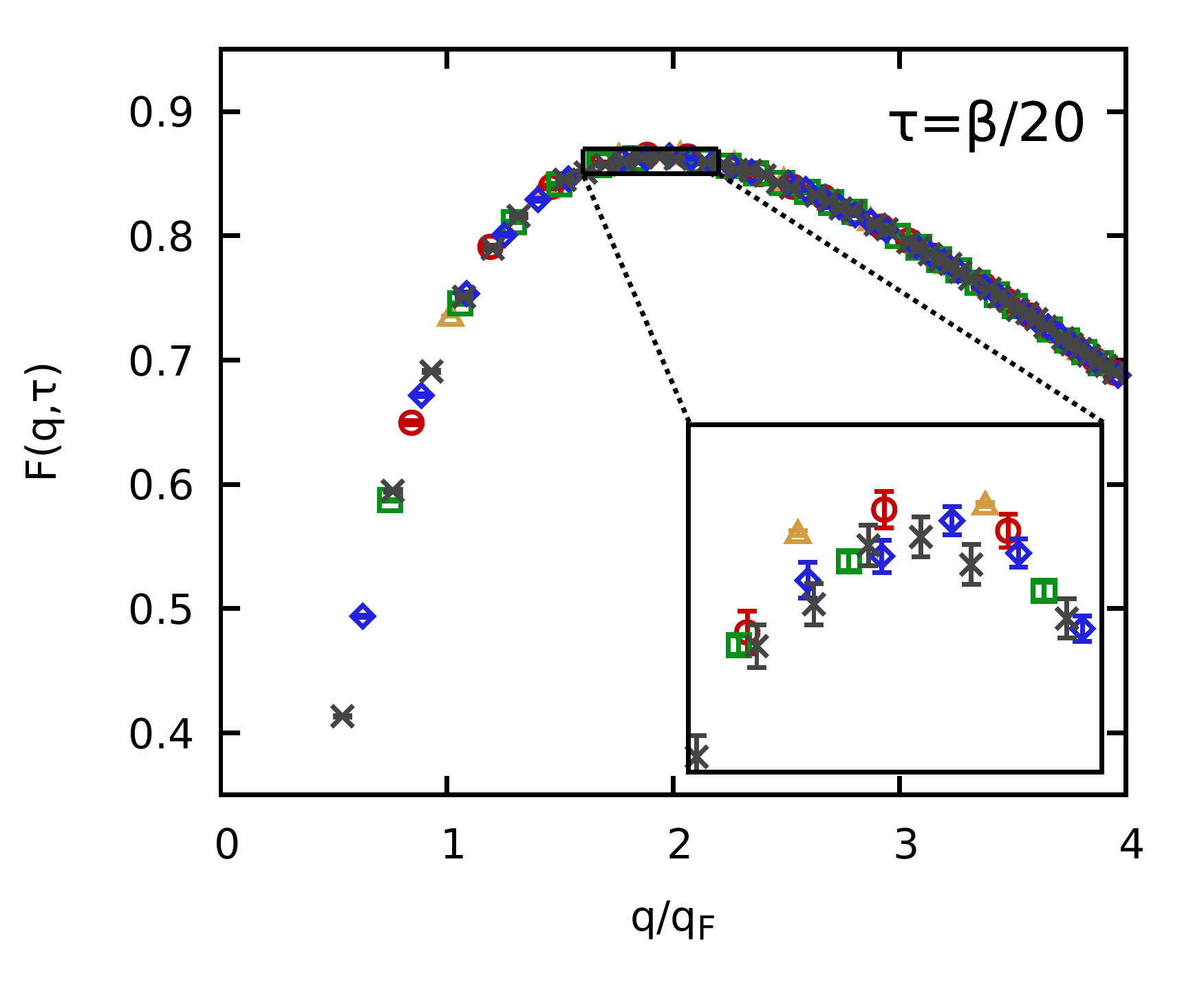}\\
\includegraphics[width=0.475\textwidth]{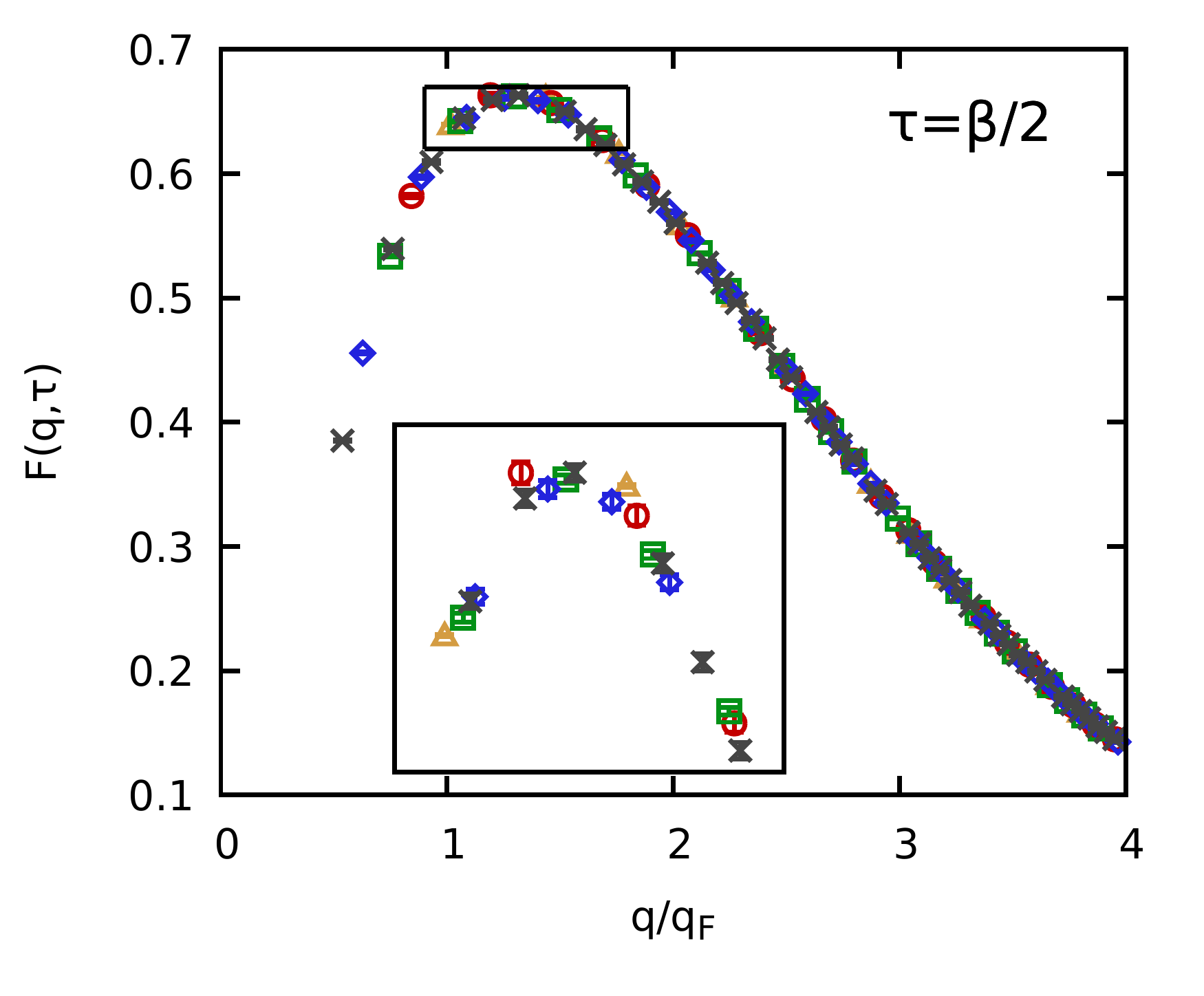}
\includegraphics[width=0.475\textwidth]{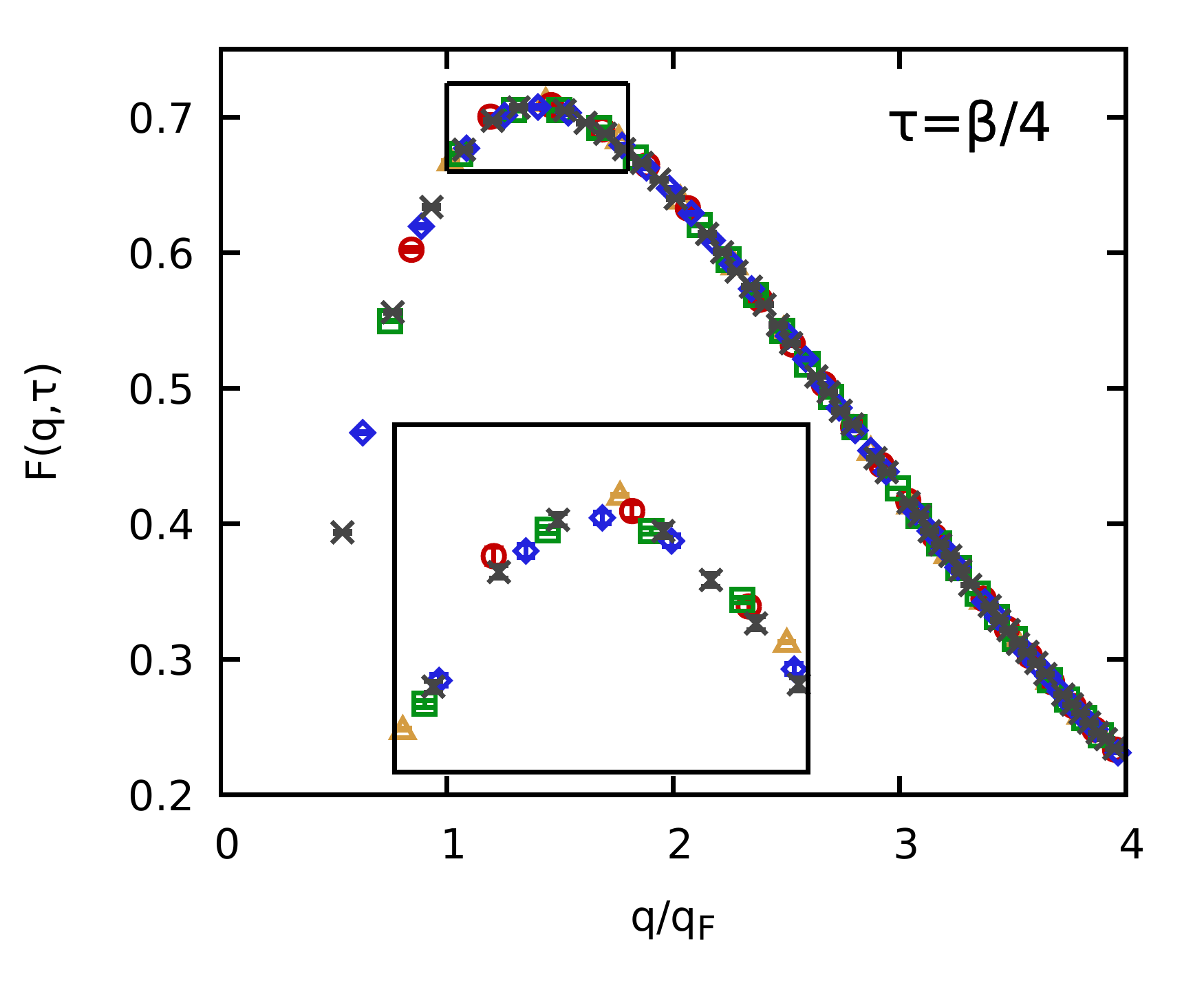}
\caption{\label{fig:FPLOT_rs2_theta2_panel}
PIMC data for the imaginary-time density--density correlation function $F$ [cf.~Eq.~(\ref{eq:define_F})] for $r_s=2$ and $\theta=2$. Shown is the $q$-dependence for different imaginary-time arguments, $\tau=0$ (i.e., $F(q,0)=S(q)$), $\tau=\beta/20$, $\tau=\beta/4$, and $\tau=\beta/2$ (clockwise). The different symbols correspond to $N=14,20,34,$ and $54$ electrons. The insets show magnified segments around the respective maxima.
}
\end{figure*}  
A more systematic investigation is presented in Fig.~\ref{fig:FPLOT_rs2_theta2_panel}, where we show $F(q,\tau)$ for four fixed values of the imaginary time along the $q$-direction. This has the advantage that we can directly compare PIMC results for different particle numbers, namely $N=8$ (yellow triangles), $N=14$ (red circles), $N=20$ (green squares), $N=34$ (blue diamonds), and $N=54$ (grey crosses). As a side note, we mention that it is fully sufficient to consider imaginary time values within the interval $\tau\in[0,\beta/2]$, as $F(q,\tau)$ is symmetric around $\beta/2$.

In the top left panel, we show results for $\tau=0$, i.e., for the usual static structure factor $S(q)$. At these conditions, there is not much spatial structure in the system, and $S(q)$ is a monotonically increasing function without any correlation-induced peaks. Moreover, the PIMC data for the different particle numbers are in excellent agreement over the entire depicted $q$-range, and no system-size dependence can be resolved within the given statistical uncertainty even for as few as $N=8$ electrons. While certainly being remarkable, this results is not unexpected, and similar observations have been reported both at finite temperature~\cite{review,dornheim_cpp,dornheim_prl} and in the ground state~\cite{Chiesa_PRL_2006,Holzmann_Finite_Size_2016}.

Moving on to $\tau=\beta/20$ (top right panel), the situation somewhat changes. Firstly, $F(q,\tau=\beta/20)$ is no longer monotonically increasing, but exhibits a maximum around twice the Fermi wave number $q_\textnormal{F}$. In addition, the results for $N=8$ are clearly shifted upward compared to the other particle numbers. A similar effect for $N=14$ is possible, but cannot be confirmed due to the given Monte-Carlo error bars, and all other data agree within the statistical uncertainty.
For $\tau=\beta/4$ (bottom left panel) and $\tau=\beta/2$ (bottom right panel), we find the same trends regarding $N$ as for $\tau=\beta/20$, and the maximum is somewhat shifted towards smaller wave numbers with increasing $\tau$.

Based on this investigation of $F(q,\tau)$ alone, we would thus predict that the reconstructed spectra for $N\geq20$ exhibit no finite-size effects, whereas they seem possible for $N=14$ and even likely for $N=20$.

\begin{figure*}\centering
\includegraphics[width=0.475\textwidth]{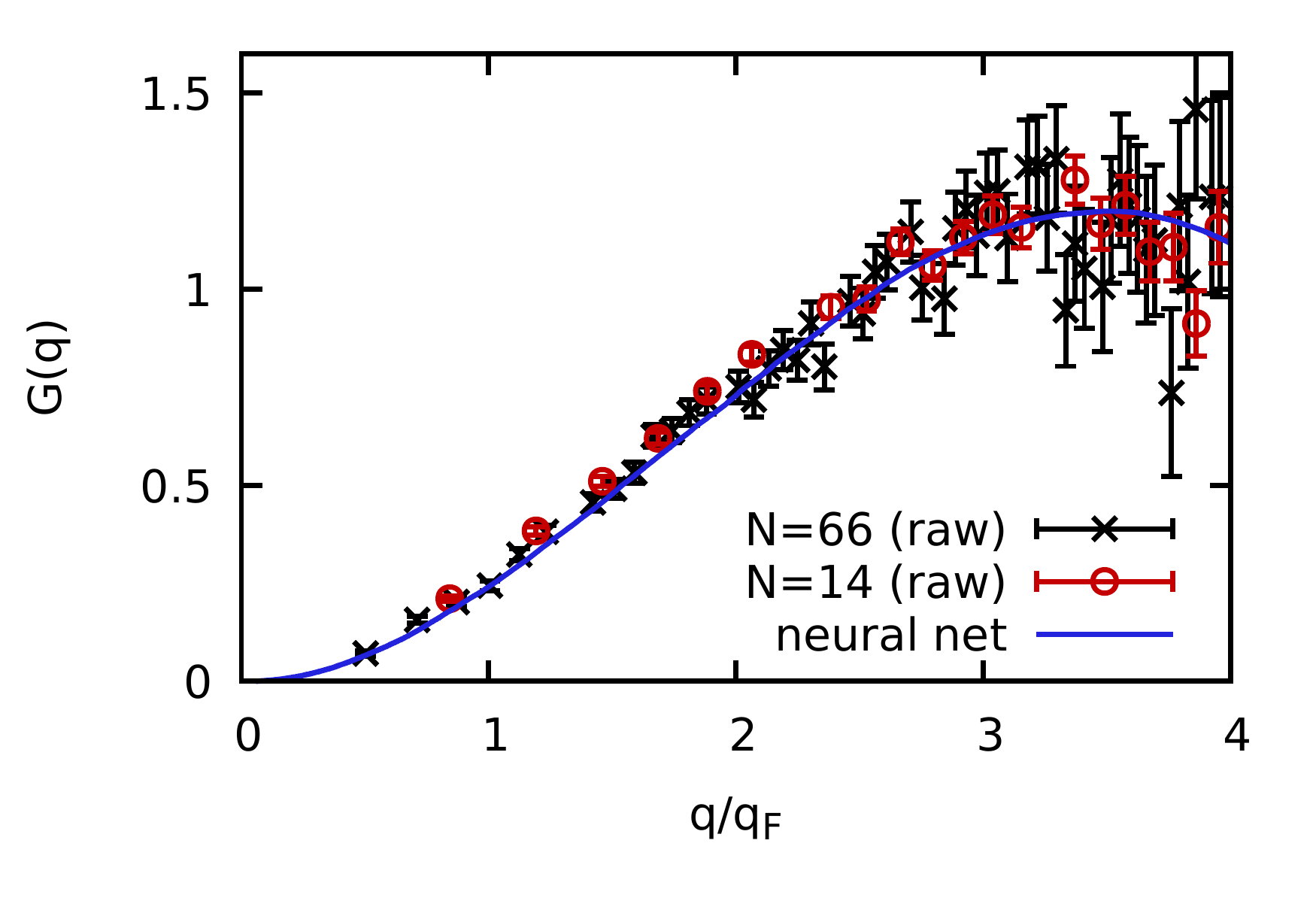}\includegraphics[width=0.475\textwidth]{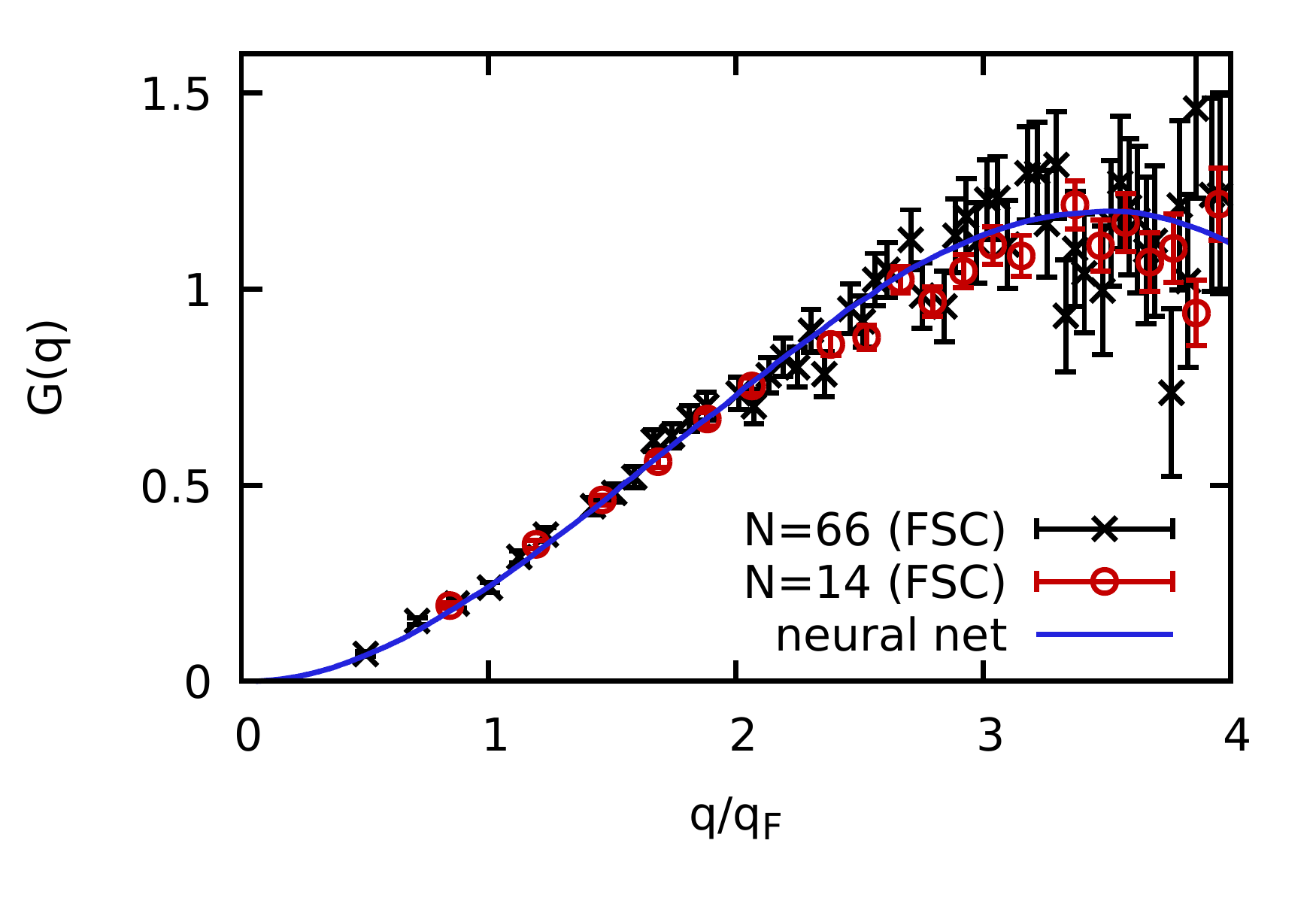}
\caption{\label{fig:G_static_rs2_theta2}
PIMC data for the $q$-dependence of the static local field correction $G(q)=G(q,0)$ for $r_s=2$ and $\theta=2+$. The black crosses and red circles depict simulation results for $N=66$ and $N=14$ electrons, and the blue curve has been obtained from the neural-net representation from Ref.~\cite{dornheim_ML}. The left and right panels corresponds to the raw and finite-size corrected (FSC) PIMC data, respectively.
}
\end{figure*}

The second main ingredient to the reconstruction is the static limit of the LFC, i.e., $G(q)=\textnormal{Re}G(q,0)$, which can be readily computed from by solving Eq.~(\ref{eq:chi}) for $G$ after inserting the PIMC data for $\chi(q)$ that is obtained via Eq.~(\ref{eq:static_chi}). This procedure is explained in detail, e.g., in Refs.~\cite{dornheim_ML,dornheim_HEDP}.
The results are shown in 
Fig.~\ref{fig:G_static_rs2_theta2}
for $N=14$ (red circles) and $N=66$ (grey crosses) electrons. Let us first focus on the left panel that has been directly obtained from the PIMC data for $\chi(q)$ without any additional finite-size correction. The solid blue line corresponds to the recent neural-net representation~\cite{dornheim_ML} of $G(q)$ and has been included as a guide to the eye. Evidently, all three curves exhibit a similar progression and can hardly be distinguished within the Monte-Carlo error bars. Yet, we note that the $N=66$ data points are in excellent agreement to the neural net, whereas the $N=14$ data appear to be somewhat too high for small wave numbers $q$.

The explanation is illustrated in the right panel, where the PIMC data for $G(q)$ have been finite-size corrected (see Ref.~\cite{groth_jcp} for a detailed discussion of the finite-size correction of both $\chi(q)$ and $G(q)$). While the grey data set is hardly affected by this procedure, the red circles are shifted downwards towards the neural net representation. We thus note that finite-size effects in the static limit of the local field correction are quite small, but are noticeable in the case of $N\leq14$ electrons.

\begin{figure}\centering
\includegraphics[width=0.475\textwidth]{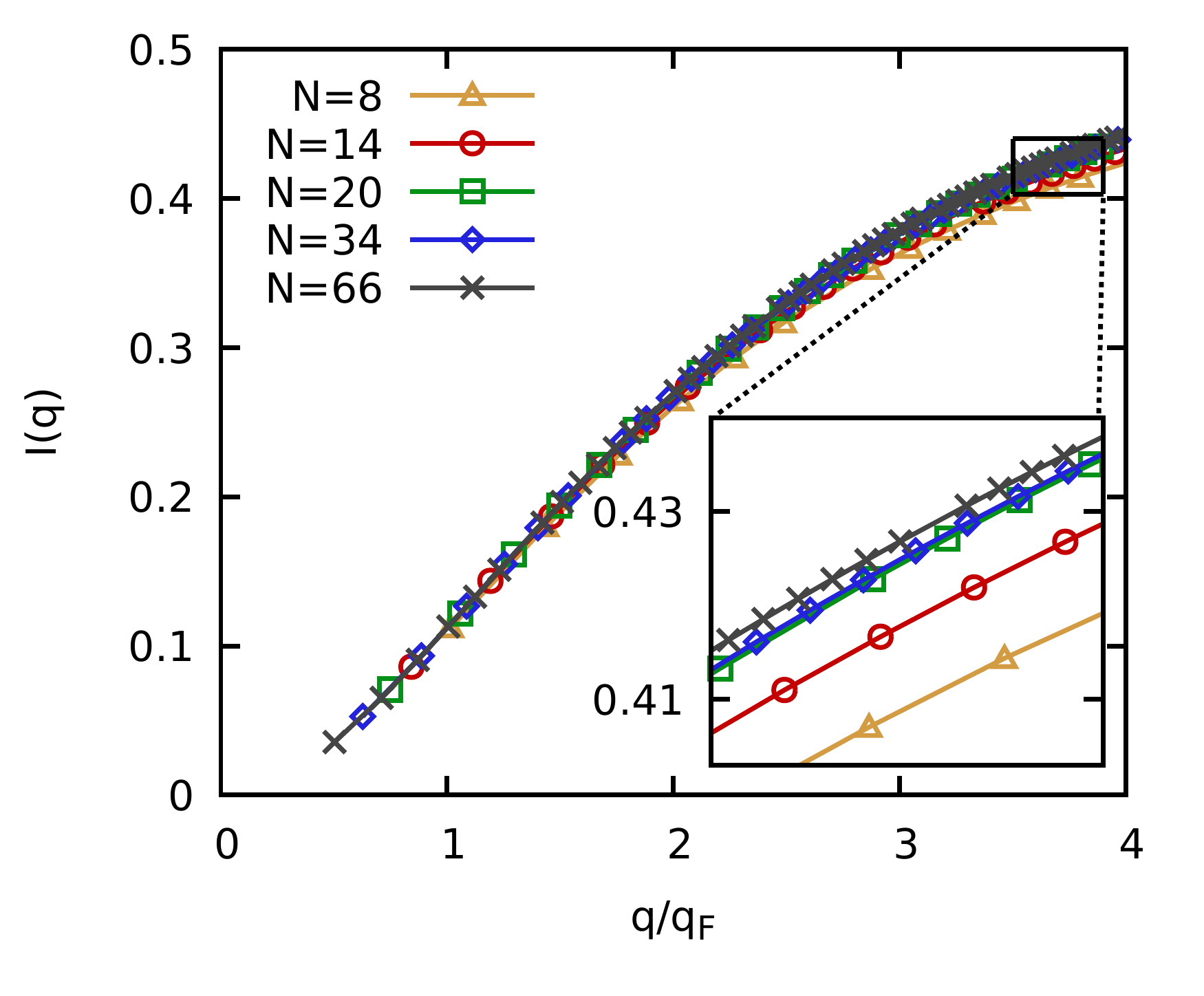}
\caption{\label{fig:Ginfty_rs2_theta2}
PIMC data for the $q$-dependence of the interaction function $I(q)$ [see Eq.~(\ref{eq:I})] for $r_s=2$ and $\theta=2$. The different symbols correspond to $N=8,14,20,34,$ and $66$ electrons and the inset shows a magnified segment for large $q$.
}
\end{figure}

Let us conclude our analysis of the system-size dependence of the ingredients to the reconstruction procedure with an investigation of the high-frequency limit $G(q,\infty)=\textnormal{Re}G(q,\infty)$, which is defined in Eq.~(\ref{eq:Ginfty}) in Sec.~\ref{sec:reconstruction}. Apart from some trivial pre-factors, this limit is defined by a) the exchange--correlation contribution to the kinetic energy $K_\textnormal{xc}$ and b) the interaction integral Eq.~(\ref{eq:I}) that depends on the static structure factor $S(q)$.
Contribution a) is computed from the parametrization of the exchange--correlation free energy (in the thermodynamic limit) by Groth \textit{et al.}~\cite{groth_prl}, and, therefore, does not depend on the particle number $N$. The evaluation of contribution b), on the other hand, is nontrivial and deserves our attention.

Evidently, Eq.~(\ref{eq:I}) requires to integrate the static structure factor $S(q)$ (or, to be precise, a function thereof) over continuous momenta $q$. This, however, is problematic as 1) PIMC data for $S(q)$ are only available for discrete $q$ values and 2) no PIMC data are available below a minimum value of $q_\textnormal{min}=2\pi/L$. In practice, we overcome this obstacle by performing cubic basis spline fits that combine the exact long-wavelength limit of $S(q)$ that is known from the perfect screening sum-rule~\cite{kugler_bounds}
\begin{eqnarray}\label{eq:S0}
S_0(q) = \lim_{q\to 0} S(q) = \frac{q^2}{2\omega_\textnormal{p}}\textnormal{coth}\left(
\frac{\beta\omega_\textnormal{p}}{2}
\right)\ ,
\end{eqnarray}with a smooth interpolation of the PIMC data elsewhere; see Ref.~\cite{dornheim_cpp} for a detailed discussion of this procedure.

In Fig.~\ref{fig:Ginfty_rs2_theta2}, we show results for the interaction integral $I(q)$ that have been obtained by integrating the basis splines for different particle numbers. The behaviour with respect to $N$
is quite similar to $F(q,\tau)$ and we find noticeable finite-size effects for both $N=14$ and $N=8$, which increase towards large wave numbers.

\begin{figure}\centering
\includegraphics[width=0.475\textwidth]{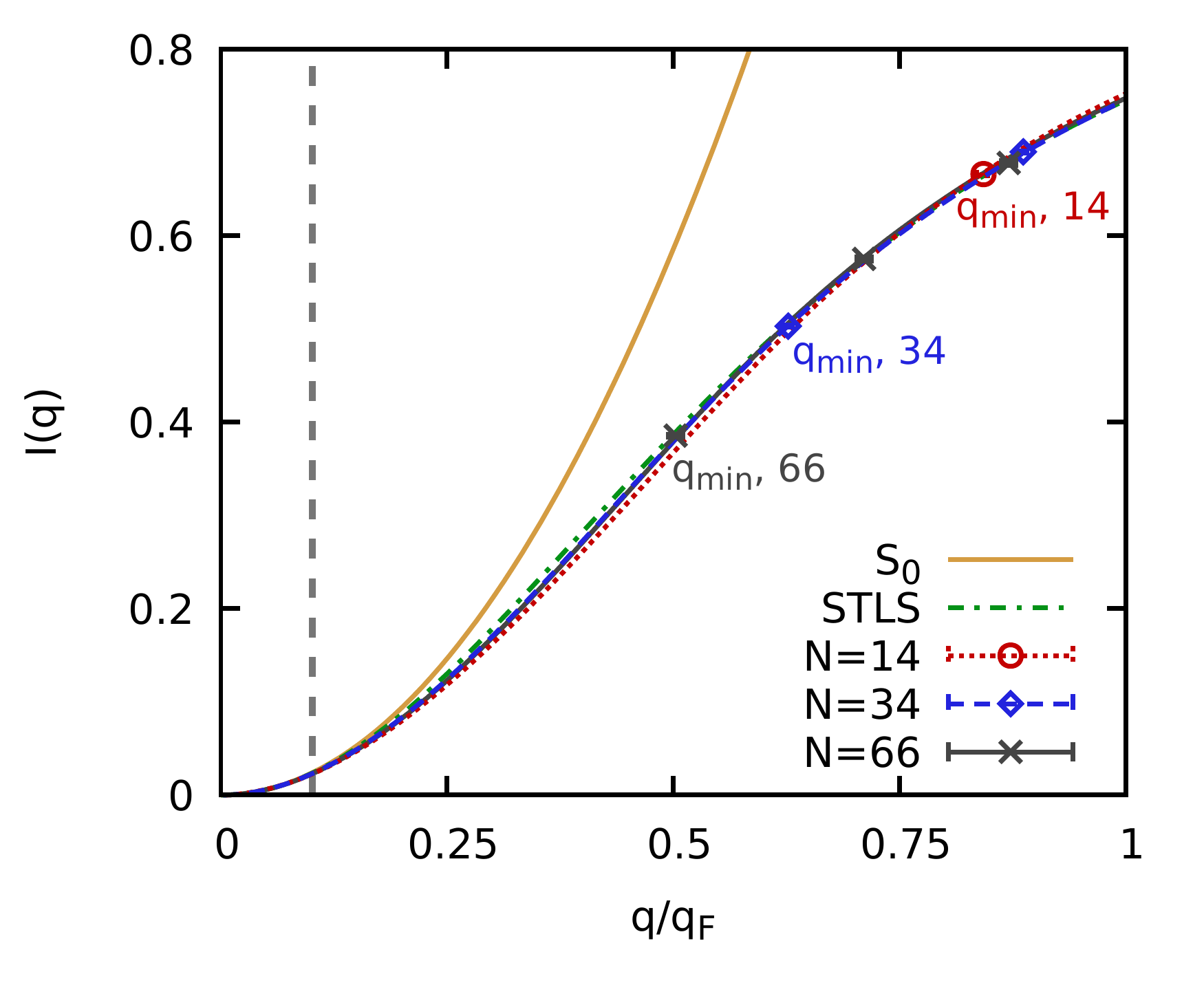}
\caption{\label{fig:BSPLINE_rs2_theta2}
Spline interpolation of the static structure factor. The solid yellow and dash-dotted green lines correspond to $S_0(q)$ [cf.~Eq.~(\ref{eq:S0})] and STLS. The red circles, blue diamonds, and dark grey crosses show PIMC results for $S(q)$ obtained for $N=14$, $N=34$, and $N=66$ electrons, respectively, and the corresponding curves cubic basis splines [see Ref.~\cite{dornheim_cpp} for details] connecting $S_0(q)$ for small $q$ with the PIMC data elsewhere. Further, the vertical dashed grey line depicts the maximum $q$-value where $S_0(q)$ is used for $N=14$.
}
\end{figure}  

This can be understood by investigation the corresponding spline-fits of the static structure factor, which is shown in Fig.~\ref{fig:BSPLINE_rs2_theta2}. The different symbols correspond to the $q$-values that are accessible in a PIMC simulation of $N=14$ (red circles), $N=34$ (blue diamonds), and $N=66$ (grey crosses) electrons, and the solid yellow curve to the exact long-wavelength behaviour of $S(q)$ that is given by Eq.~(\ref{eq:S0}). The corresponding spline representations of $S(q)$ have been obtained by combining the yellow curve up to an empirically chosen maximum $q$-value (cf.~the vertical dashed grey line) with the PIMC data for a specific $N$ elsewhere. 
The dash-dotted green curve corresponds to the finite-temperature version~\cite{stls,stls2} of the dielectric theory by Singwi \textit{et al.}~\cite{stls_original}
and has been included as a reference.
Evidently, the splines for $N=34$ and $N=66$ electrons cannot be distinguished over the entire depicted $q$-range, and the blue curve correctly predicts the data point for $S(q_\textnormal{min})$ for $N=66$ although it lies outside the fitting range. For $N=14$ (dotted red line), on the other hand, the range between $q_\textnormal{min}$ and the validity range of $S_0(q)$ is significantly larger, which makes the interpolation in between much less reliable. As a result, the red curve substantially deviates from the other two for $q\sim0.5q_\textnormal{F}$, and this trend gets only more pronounced for $N=8$.
The subsequent integration over these splines in Eq.~(\ref{eq:I}) then leads to the finite-size effects in $I(q)$ observed in Fig.~\ref{fig:Ginfty_rs2_theta2}.

In a nutshell, we have found that the input data for the reconstruction procedure are afflicted with substantial finite-size errors for $N=8$ and noticeable errors for $N=14$ electrons.

\begin{figure*}\centering\includegraphics[width=0.48\textwidth]{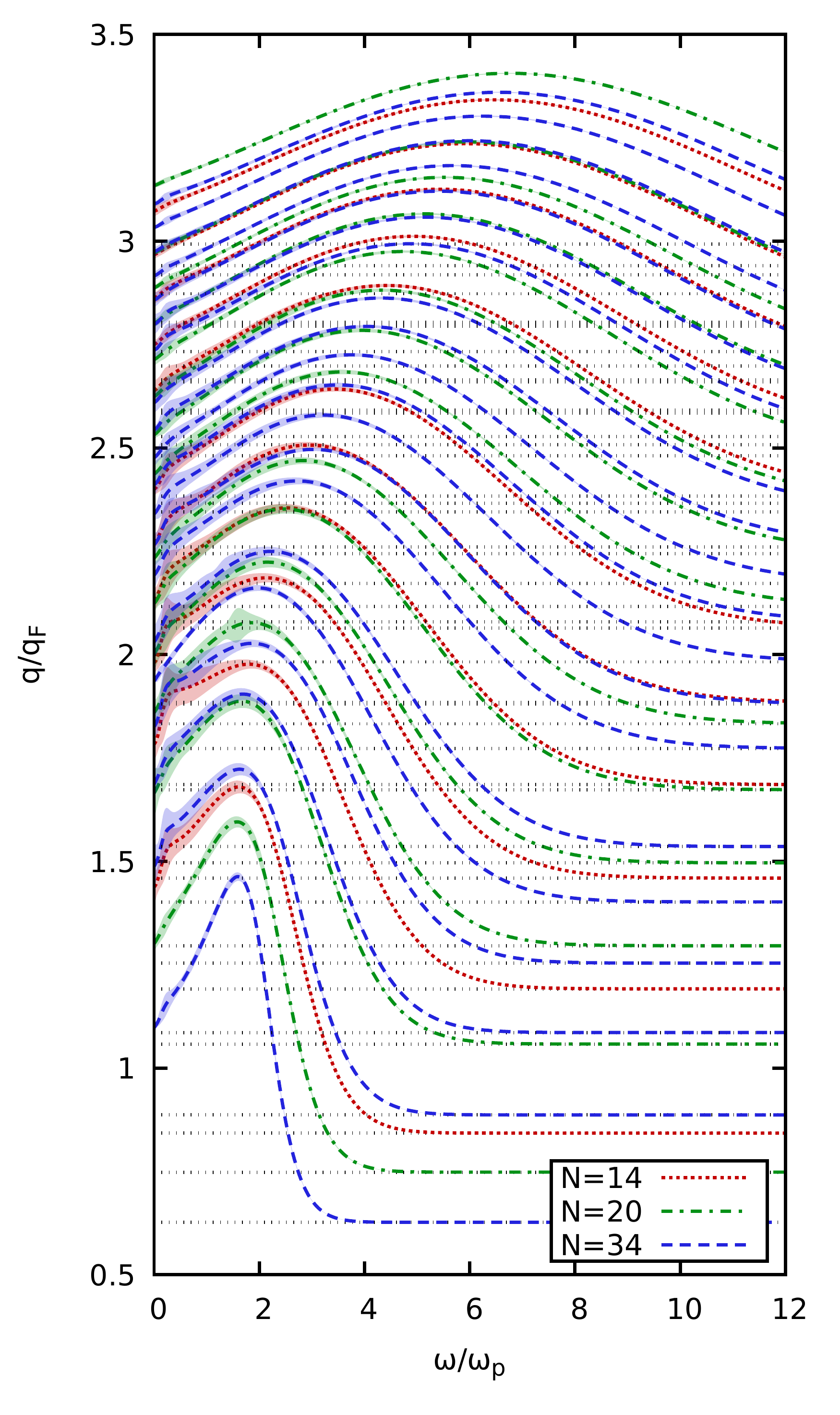}\includegraphics[width=0.48\textwidth]{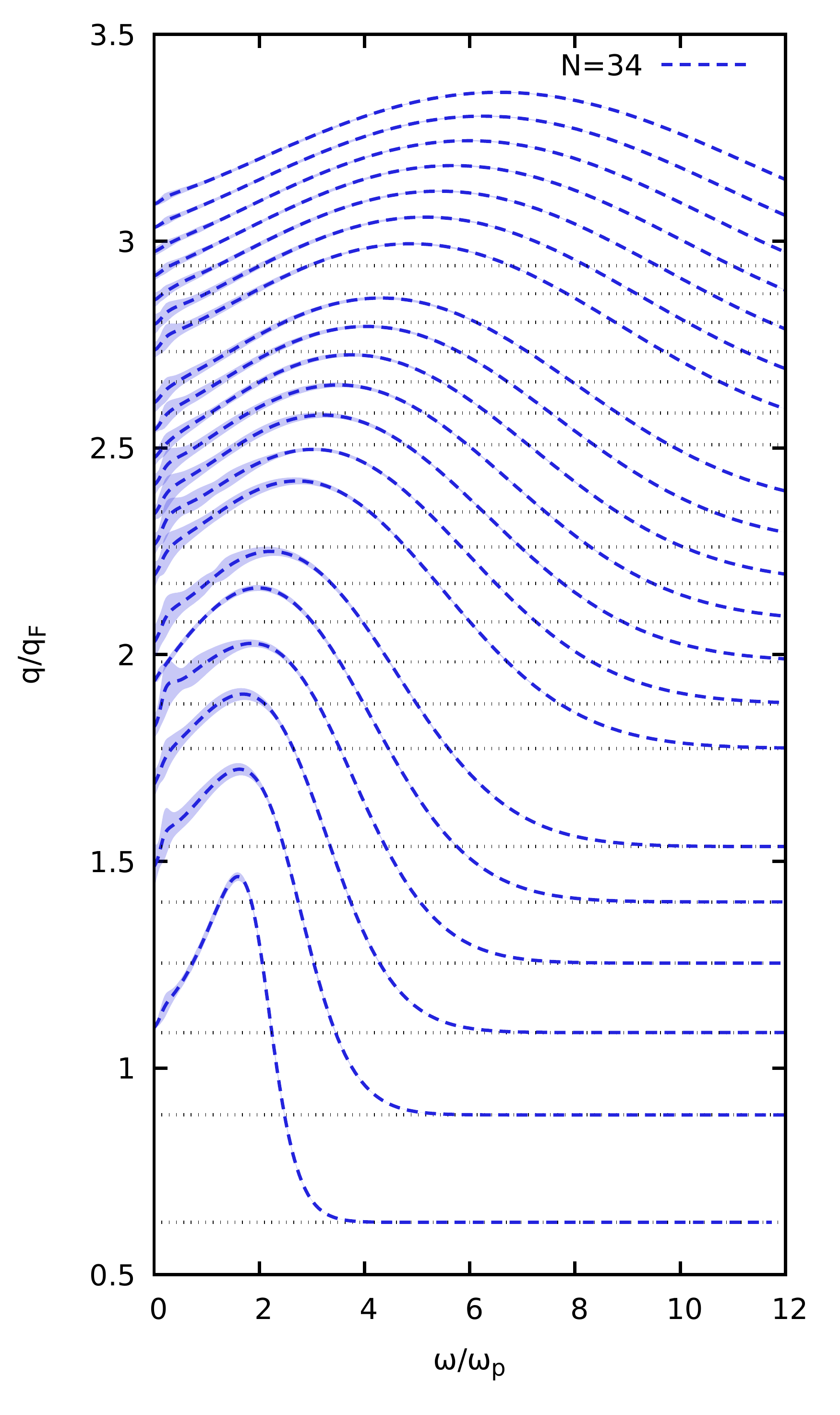}
\caption{\label{fig:Spectral_DSF_rs2_theta2_COMPLETE}
Reconstructed dynamic structure factors at WDM conditions ($r_s=2$ and $\theta=2$). Left panel: Solutions for $N=14$ (dotted red), $N=20$ (dash-dotted green), and $N=34$ (dashed blue) electrons at selected $q$-values. Right panel: Solutions for all $q$-values in the depicted $q$-range for $N=34$.
}
\end{figure*}

This brings us to the central topic of this work, which is the investigation of the system-size effects in the reconstructed dynamic structure factors $S(q,\omega)$
themselves. To this end, we show the full frequency-dependence of $S(q,\omega)$ for selected wave numbers in the left panel of Fig.\ref{fig:Spectral_DSF_rs2_theta2_COMPLETE} for $N=14$ (red dotted), $N=20$ (dash-dotted green), and $N=34$ (dashed blue) electrons. For completeness, we note that it is sufficient to only show the positive $\omega$-range, since the DSF obeys the detailed balance relation~\cite{quantum_theory}
\begin{eqnarray}
S(q,-\omega) = S(q,\omega) e^{-\beta\omega}\ .
\end{eqnarray}

Let us start by briefly touching upon the physical effects. In the limit of small wave numbers, the random phase approximation becomes exact and $S(q,\omega)$ becomes a delta-function around the plasma frequency $\omega_\textnormal{p}=\sqrt{3/r_s^3}$. With increasing $q$, the DSF broadens in the frequency domain, and the normalization [i.e., the static structure factor $S(q)$, cf.~Eq.~(\ref{eq:Sq})] increases until it saturates around one.
Here, too, the random phase approximation becomes exact, as the impact of the local field correction is reduced by the $4\pi/q^2$ pre-factor, cf.~Eq.~(\ref{eq:chi}).

Regarding the reconstructed solutions for $S(q,\omega)$ for different $S(q,\omega)$, we find that the subsequent curves exhibit a smooth progression in the $q$-$\omega$-plane, and even the results for $N=14$ electrons do not exhibit any noticeable deviations from this trend.
While only certain wave numbers are included in this plot, we show the full dispersion relation of all $q$-values in the depicted wave number range for $N=34$ electrons in the right panel of Fig.~\ref{fig:Spectral_DSF_rs2_theta2_COMPLETE}.
A common feature of both panels is given by the increased uncertainty for small frequencies, which is consistent with the previous findings from Refs.~\cite{dornheim_dynamic,dynamic_folgepaper}
in this regime. In principle, this would allow for the possibility of a diffusive feature for small $\omega$ (see, e.g., Ref.~\cite{Tkachenko_PRL_2012}), but this is not expected for the present case of a quantum one-component system.

\begin{figure*}\centering\includegraphics[width=0.475\textwidth]{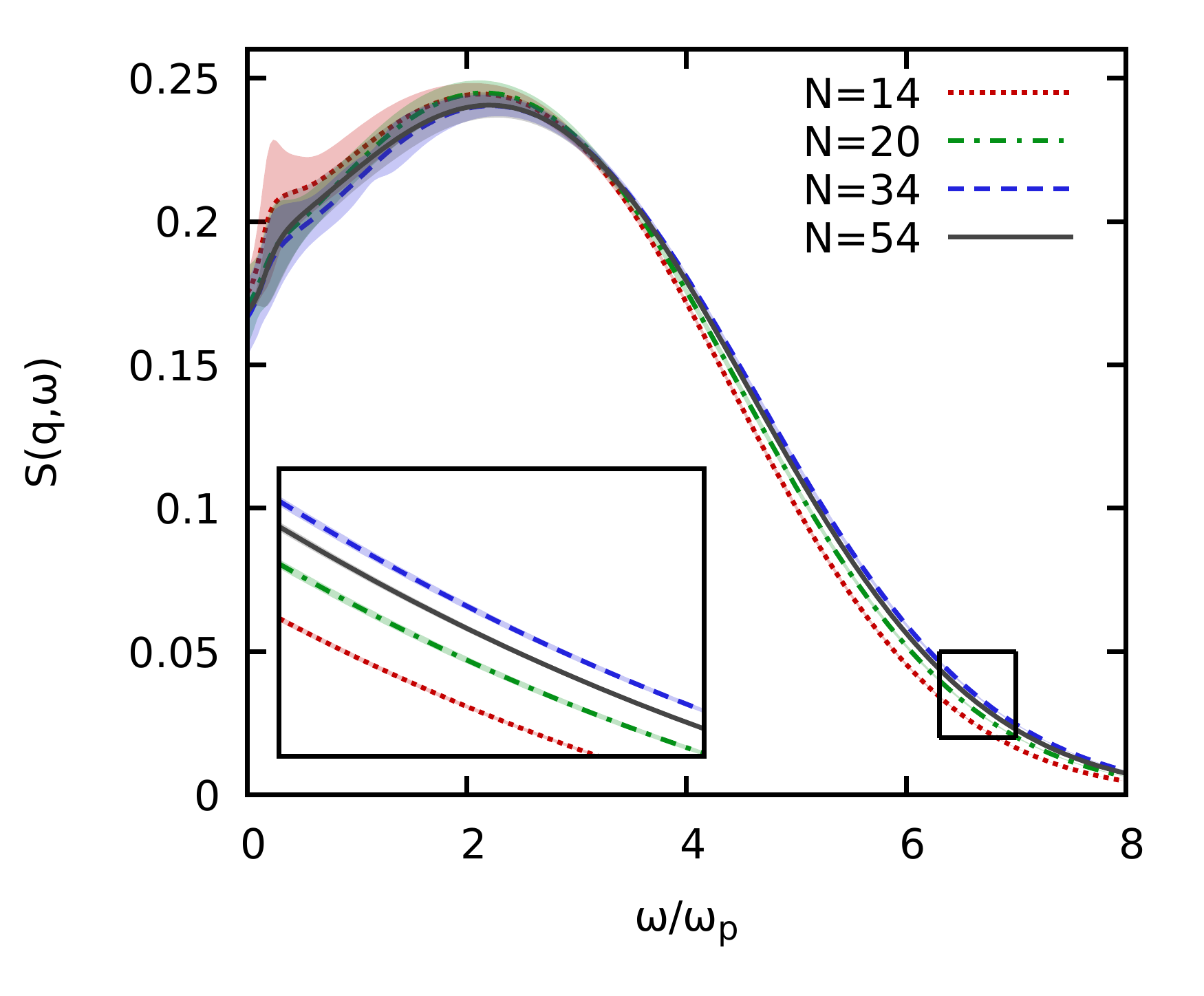}
\includegraphics[width=0.475\textwidth]{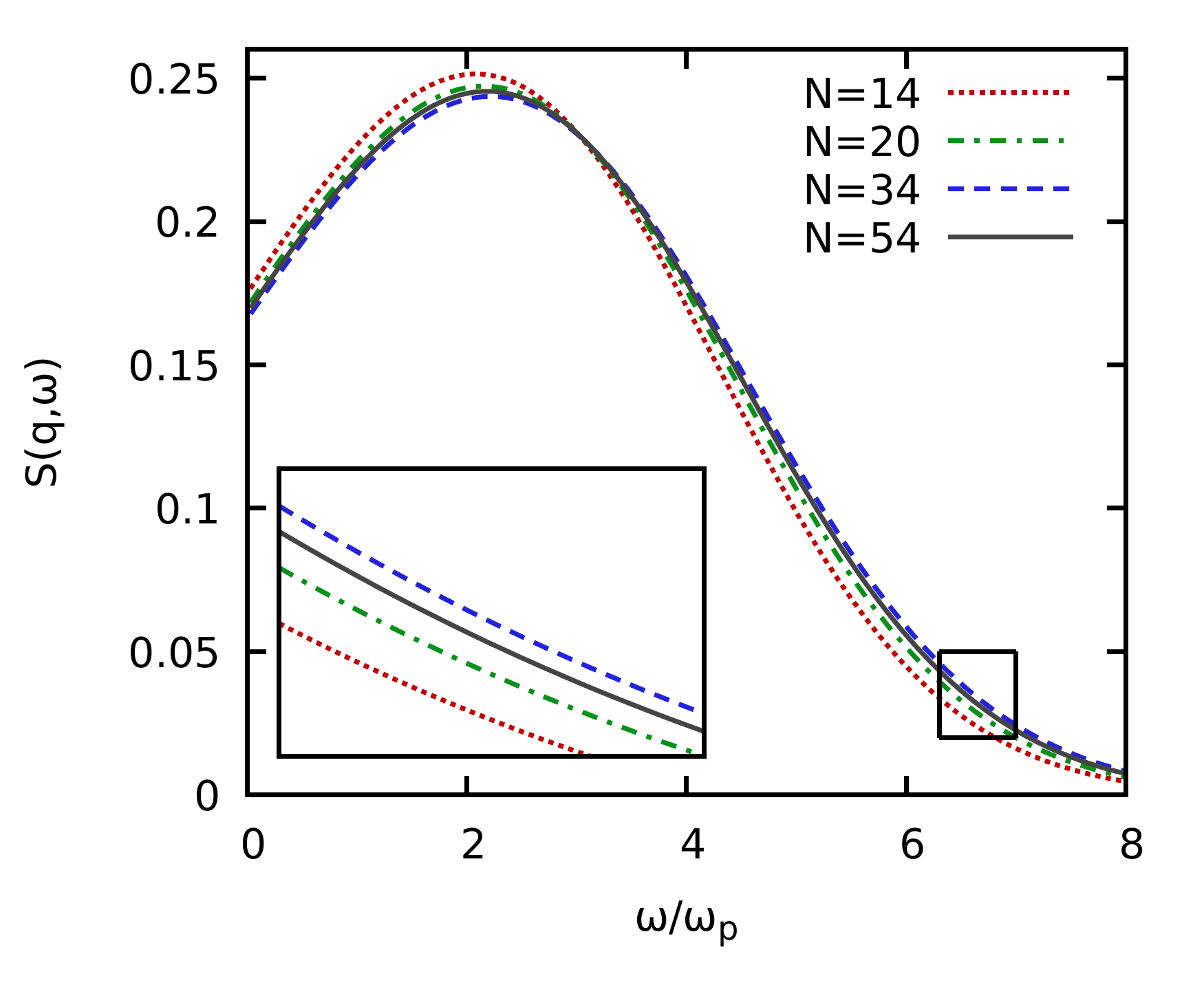}
\caption{\label{fig:DSF_rs2_theta2_multi_N}
Dynamic structure factor of the UEG at $r_s=2$ and $\theta=2$. The dotted red, dash-dotted green, dashed blue and solid black lines correspond to $N=14$ ($q/q_\textnormal{F}\approx1.460$), $N=20$ ($q/q_\textnormal{F}\approx1.496$), $N=34$ ($q/q_\textnormal{F}\approx1.536$), and $N=54$ ($q/q_\textnormal{F}\approx1.520$) electrons, where the numbers in brackets indicate the respective wave number $q$. The left panel shows the full reconstructed solutions for $S(q,\omega)$, and the right panel the corresponding curves from the static approximation that has been included as a reference.
}
\end{figure*}

In order to more rigorously discard the possibility of finite-size effects in our data for the DSF, we show $S(q,\omega)$ for similar wave numbers around $q=1.5q_\textnormal{F}$ in the left panel of Fig.~\ref{fig:DSF_rs2_theta2_multi_N} for $N=14$ (dotted red), $N=20$ (dash-dotted green), $N=34$ (dashed blue), and $N=54$ (solid grey). First and foremost, we note that all curves exhibit a very similar behaviour over the entire $\omega$-range. Yet, while the curves agree within the given confidence interval for $\omega\lesssim3\omega_\textnormal{p}$, there appear systematic deviations for larger frequencies. 
In principle, these could be 1) due to finite-size effects, 2) inconsistencies in the reconstruction procedure that are not accounted for by the confidence interval [cf.~Eq.~\ref{eq:S_error})], or 3) due to the different $q$-values for the four cases, see the caption.

To resolve this issue, we compute the same curves within the so-called \emph{static approximation}, i.e., by inserting the exact static limit $G(q,0)$ into Eq.~(\ref{eq:chi}) to get $\chi_\textnormal{static}(q,\omega)$ and subsequently evoking the fluctuation--dissipation theorem [Eq.~(\ref{eq:FDT})] to compute the corresponding DSF $S_\textnormal{static}(q,\omega)$. The results are shown in the right panel of Fig.~\ref{fig:DSF_rs2_theta2_multi_N}, and exhibit precisely the same order as the fully reconstructed curves in the left panel.
For instance, the insets of both panels depict magnified segments for the large-$\omega$ regime, and, the curves are ordered with decreasing wave number starting from the top.

We thus conclude that the observed differences are due to explanation 3), and no finite-size effects can be resolved in the reconstructed solution for $S(q,\omega)$ even for as few as $N=14$ electrons.

In contrast, for $N=8$ our reconstruction procedure was not able to find viable solutions for $S(q,\omega)$ which were then in agreement to the PIMC data for $F(q,\tau)$ and $\braket{\omega^k}$. This means that finite-size effects do not manifest as an $N$-dependence in the reconstructed spectra, but as an inconsistency between the exact constraints on $G(q,\omega)$ (cf.~Sec.~\ref{sec:reconstruction}) and the potentially system-size dependent PIMC data.

\begin{figure}\centering
\includegraphics[width=0.55\textwidth]{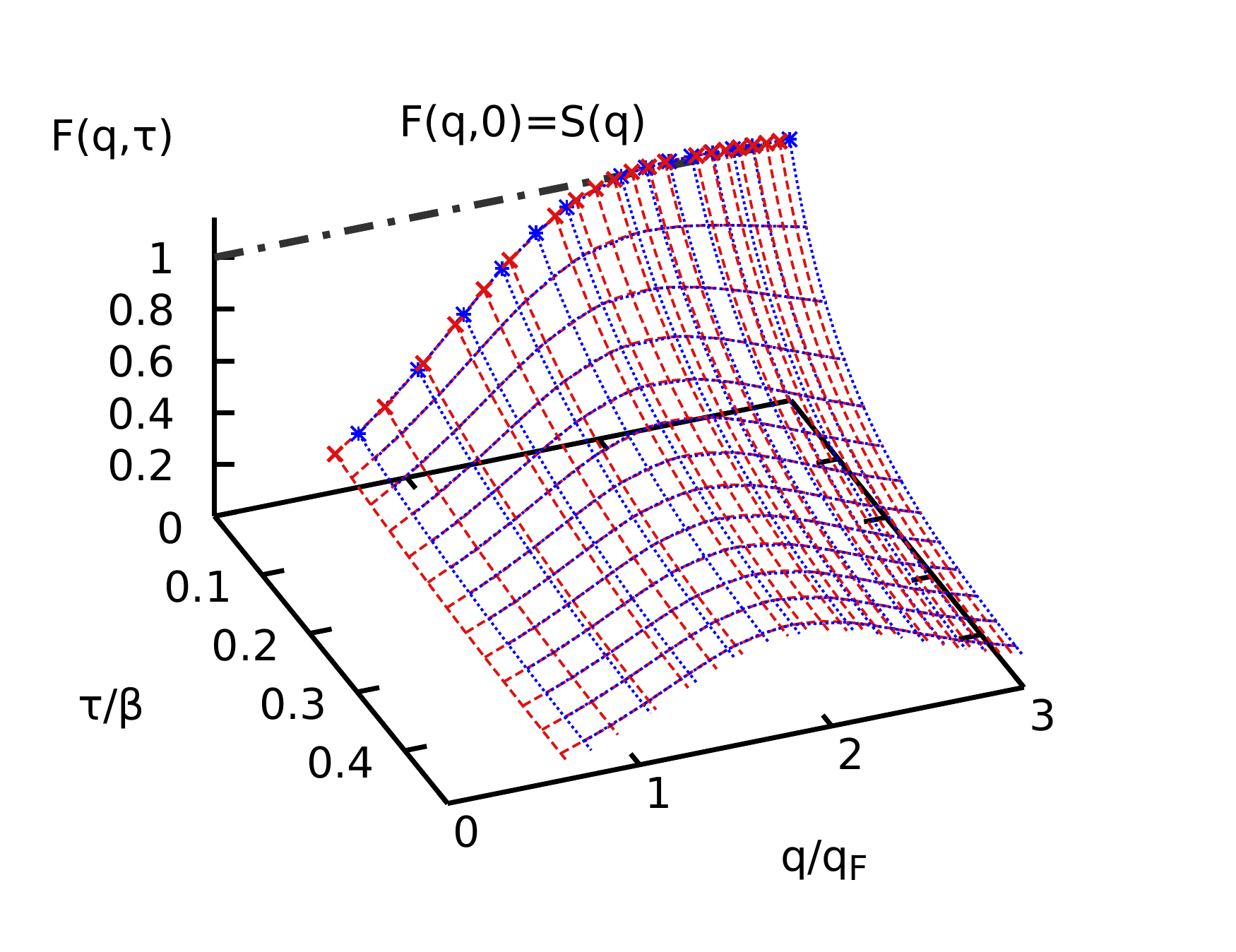}\vspace*{-0.7cm}
\caption{\label{fig:FPLOT_rs10_theta1}
PIMC data for the imaginary-time density--density correlation function $F$ [cf.~Eq.~(\ref{eq:define_F})] in the $\tau$-$x$-plane for $r_s=10$ and $\theta=1$. The red (dashed) and blue (dotted) data have been obtained for $N=34$ and $N=20$ electrons. Note that $F$ approaches the static structure factor in the $\tau=0$ limit (crosses and stars) and that it is symmetric with respect to $\tau=\beta/2$, i.e., $F(q,\tau)=F(q,\beta-\tau)$.
}
\end{figure}

\subsection{Strongly correlated regime: $r_s=10$ and $\theta=1$\label{sec:EL}}

\begin{figure*}\centering
\includegraphics[width=0.475\textwidth]{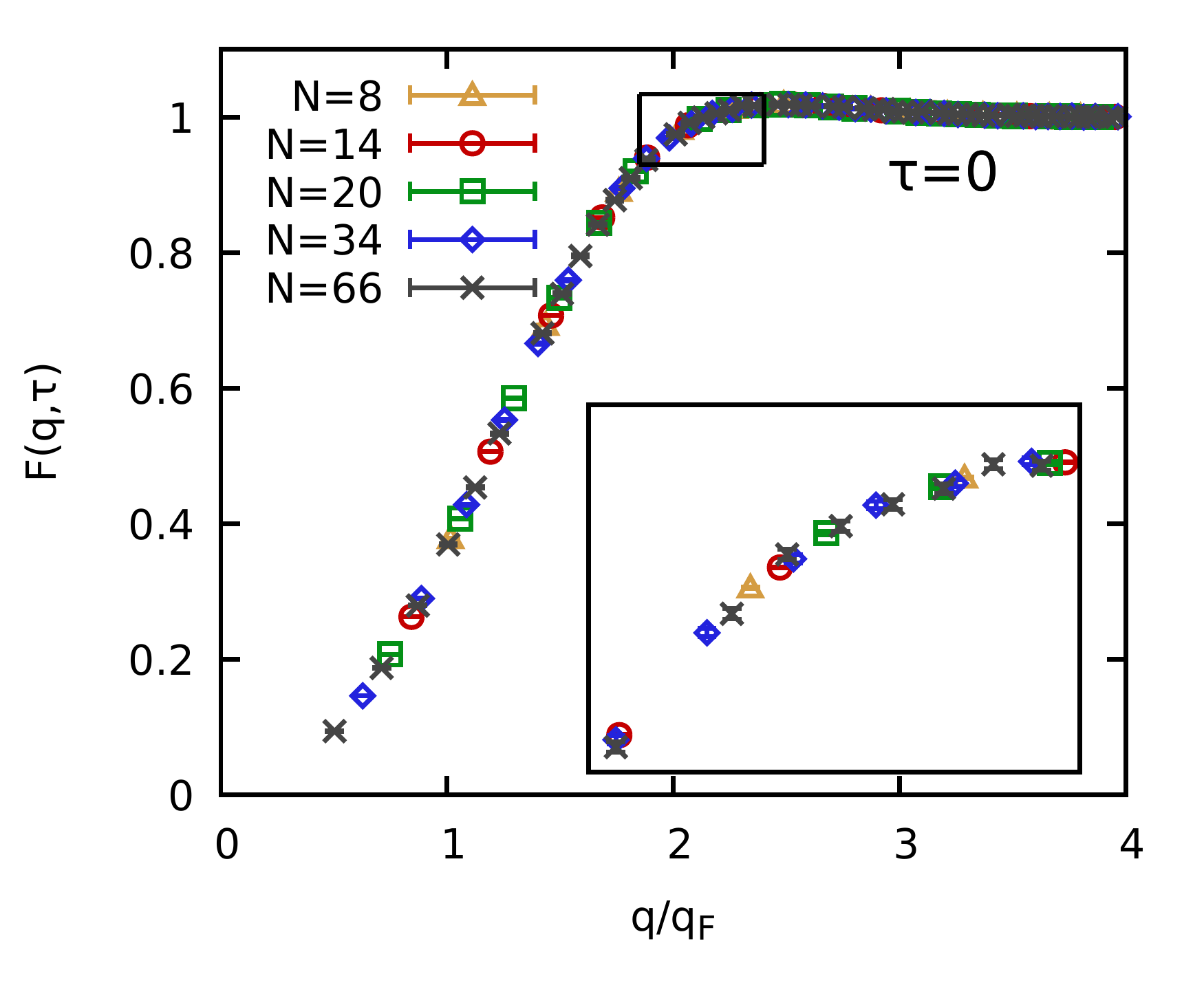}
\includegraphics[width=0.475\textwidth]{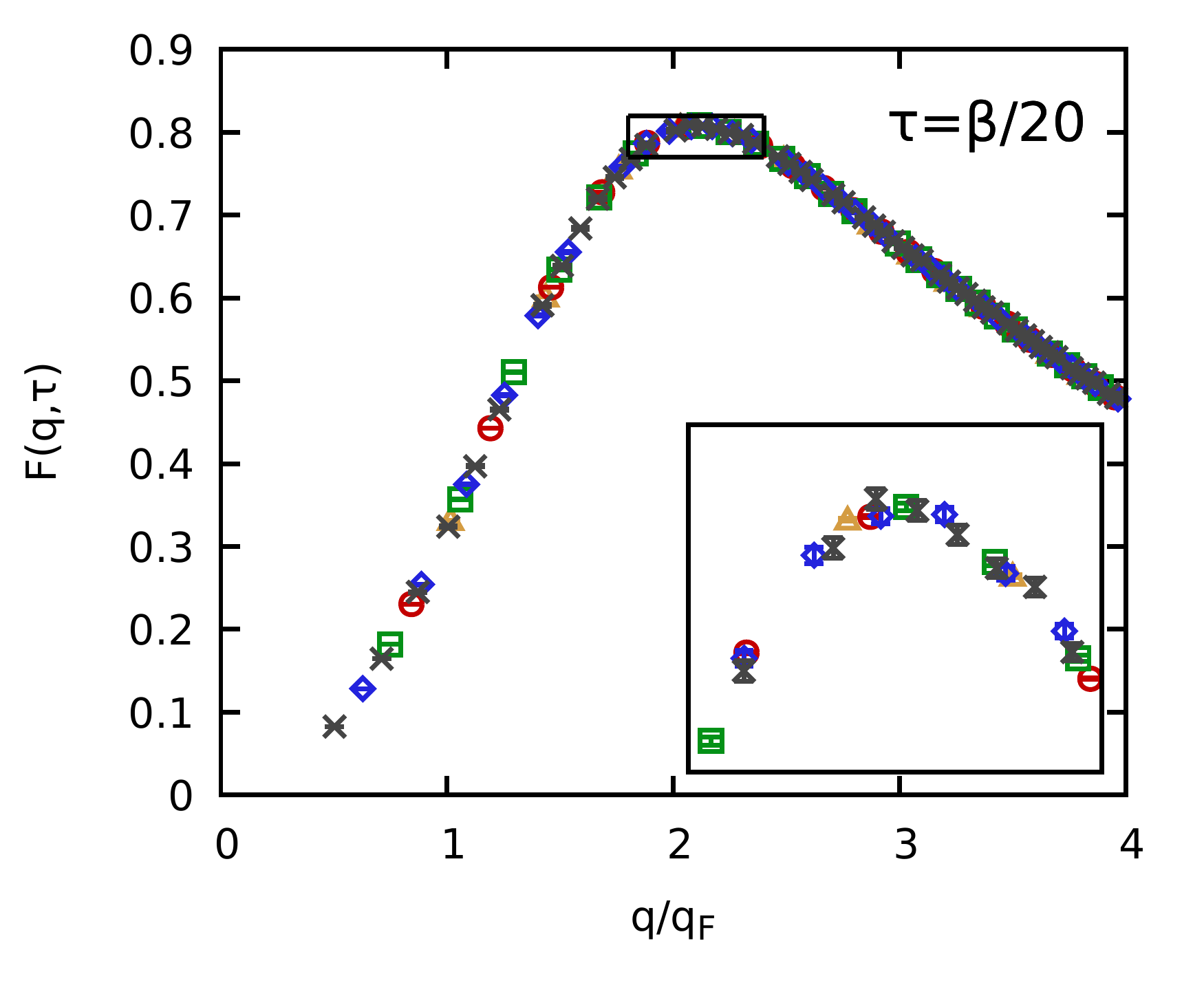}\\
\includegraphics[width=0.475\textwidth]{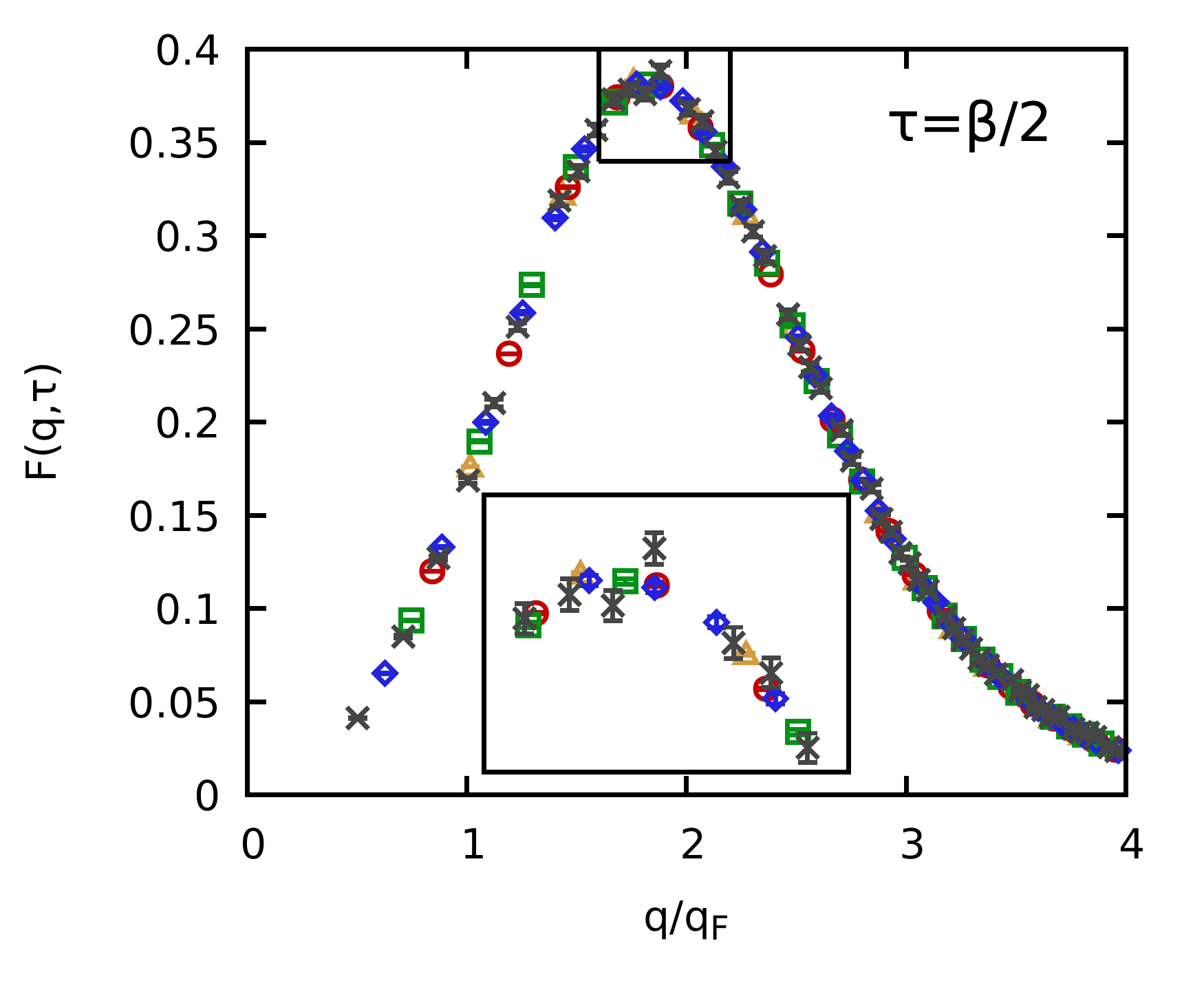}
\includegraphics[width=0.475\textwidth]{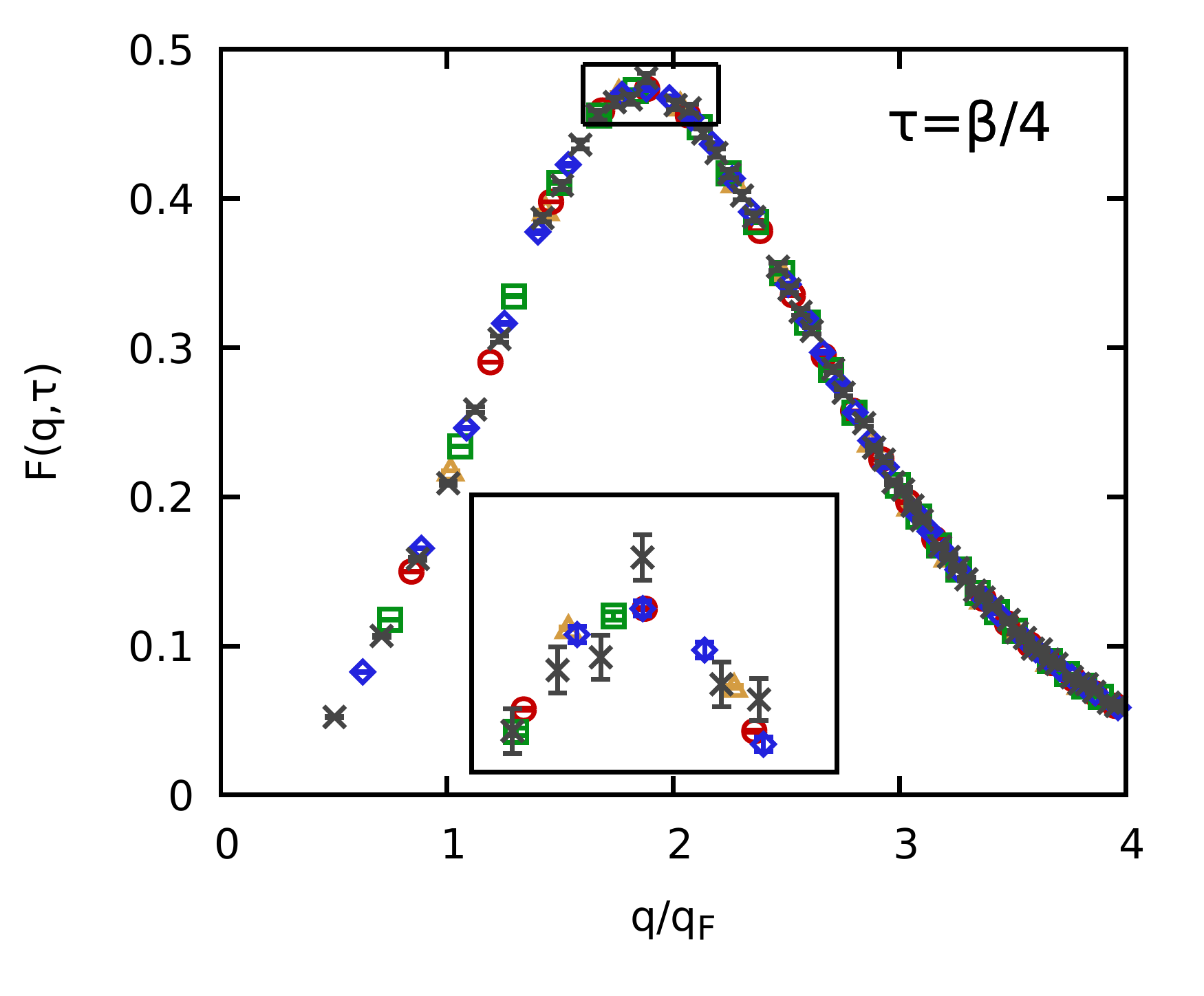}
\caption{\label{fig:FPLOT_rs10_theta1_l}
PIMC data for the imaginary-time density--density correlation function $F$ [cf.~Eq.~(\ref{eq:define_F})] for $r_s=10$ and $\theta=1$. Shown is the $q$-dependence for different imaginary-time arguments, $\tau=0$ (i.e., $F(q,0)=S(q)$), $\tau=\beta/20$, $\tau=\beta/4$, and $\tau=\beta/2$ (clockwise). The different symbols correspond to $N=8,14,20,34,$ and $66$ electrons. The insets show magnified segments around the respective maxima.
}
\end{figure*}

The second parameter regime to be explored in this work is given by the margins of the electron liquid ($r_s=10$ and $\theta=1$). Despite being less relevant for current WDM experiments, these conditions offer a plethora of interesting physical effects. Of particular relevance is a possible incipient excitonic mode that was predicted by Takada~\cite{Takada_PRB_2016} (see also Ref.~\cite{Higuchi_Japan_2000} for a discussion of the excitonic nature of this feature) and substantiated by Dornheim \textit{et al.}~\cite{dornheim_dynamic}.
Further, we mention that this regime is particularly interesting from a theoretical perspective, as the full frequency-dependence of $G(q,\omega)$ is needed for an adequate description. This is in stark contrast to the WDM regime, where using the static limit $G(q,0)$ is often sufficient to obtain highly accurate results for all $\omega$.

Since this analysis is mostly analogous to the discussion of WDM parameters in the previous section, here we only briefly state the most important findings. In Fig.~\ref{fig:FPLOT_rs10_theta1}, we show the imaginary-time density--density correlation function in the $q$-$\tau$-plane again for $N=20$ (blue) and $N=34$ (red) unpolarized electrons. In contrast to the WDM example shown in Fig.~\ref{fig:FPLOT_rs2_theta2}, here $F(q,\tau)$ exhibits a more complicated structure, and the static structure factor $S(q)=F(q,0)$ has a small maximum around twice the Fermi wave number and is thus non monotonous. Although, in general, the direct physical interpretation of this quantity is rather difficult, it was found that the amount of structure substantially increases with coupling strength, with a progression of several maxima and minima in the electron liquid regime~\cite{dornheim_electron_liquid}. Still, no difference between the two system sizes can be spotted in Fig.~\ref{fig:FPLOT_rs10_theta1}.

A more detailed investigation is presented in Fig.~\ref{fig:FPLOT_rs10_theta1_l}, where we show the $q$-dependence of $F$ for fixed $\tau$-values. Interestingly, no finite size effects are evident anywhere even for as few as $N=8$ electrons. This is not fully unexpected, as the system size dependence is known to increase both with density and with temperature~\cite{dornheim_prl} at these conditions.
For completeness, we mention that for even lower temperatures ($\theta\lesssim0.25$), shell-filling effects in momentum-space become important that can be mitigated by simulating commensurate particle numbers (i.e., $N=33$ or $N=66$ electrons for a spin-polarized or unpolarized system) and by carrying out an additional twist-averaging procedure~\cite{Lin_PRE_2001,Spink_PRB_2013}.

\begin{figure*}\centering
\includegraphics[width=0.475\textwidth]{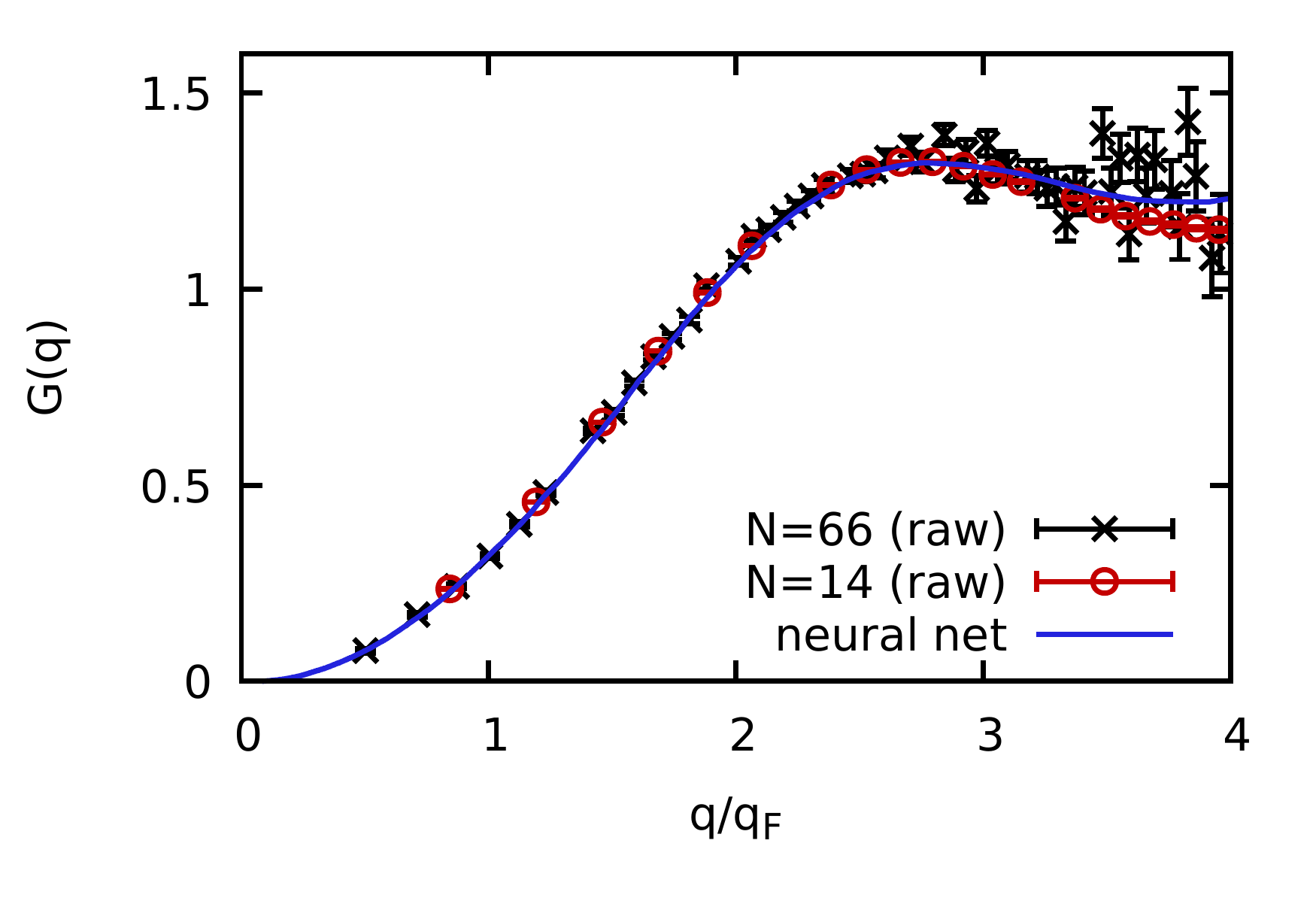}\includegraphics[width=0.475\textwidth]{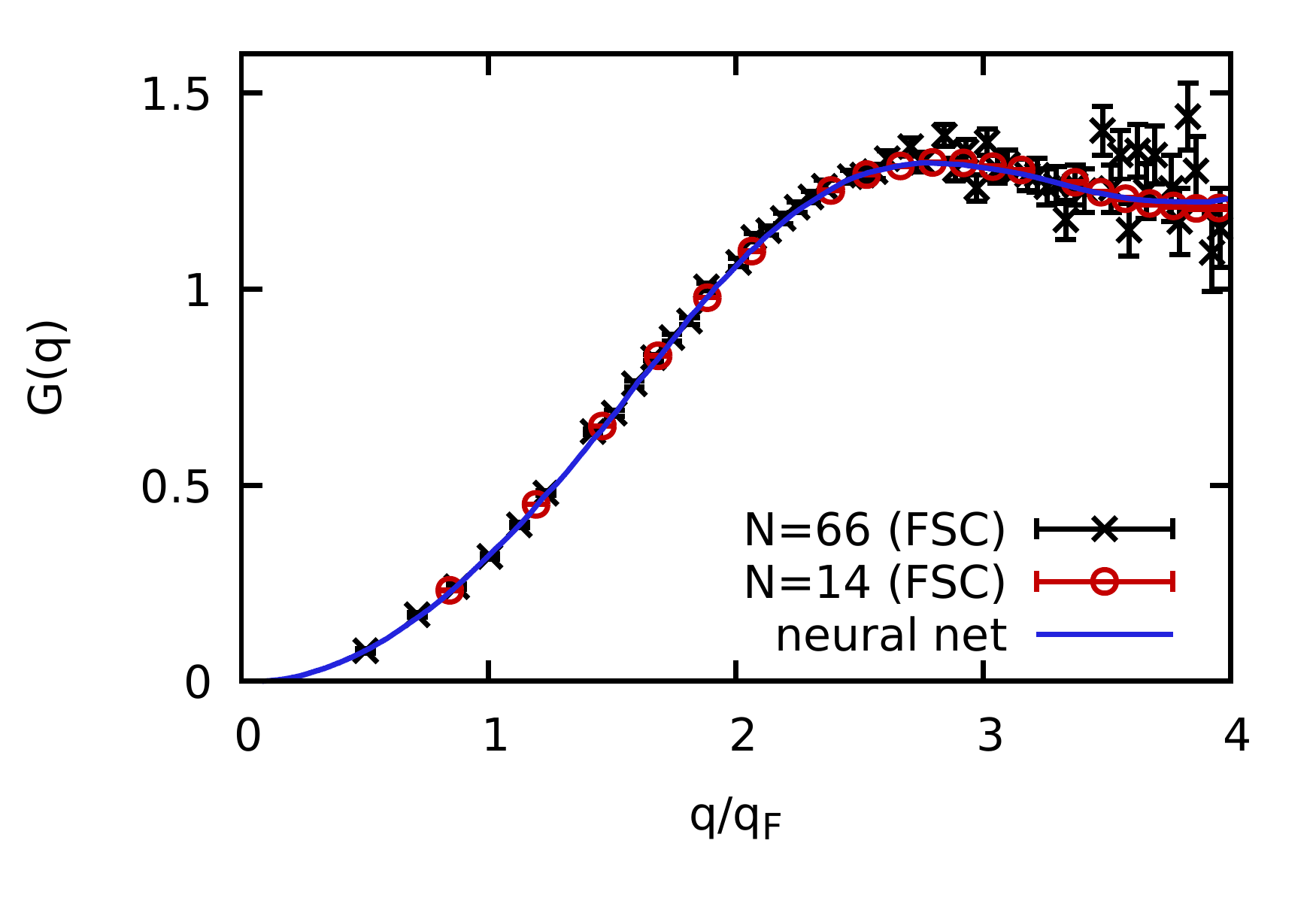}
\caption{\label{fig:G_static_rs10_theta1}
PIMC data for the $q$-dependence of the static local field correction $G(q)=G(q,0)$ for $r_s=10$ and $\theta=1$. The black crosses and red circles depict simulation results for $N=66$ and $N=14$ electrons, and the blue curve has been obtained from the neural-net representation from Ref.~\cite{dornheim_ML}. The left and right panels corresponds to the raw and finite-size corrected (FSC) PIMC data, respectively.
}
\end{figure*}

The next relevant input quantity to the reconstruction procedure is given by the static limit of $G(q,\omega)$, which is shown in Fig.~\ref{fig:G_static_rs10_theta1}. Again, even for $N=14$ electrons almost no finite-size effects are visible in the pure PIMC data (left panel), and the finite-size correction (right panel) only affects the data for $q\gtrsim3q_\textnormal{F}$.

\begin{figure}\centering
\includegraphics[width=0.475\textwidth]{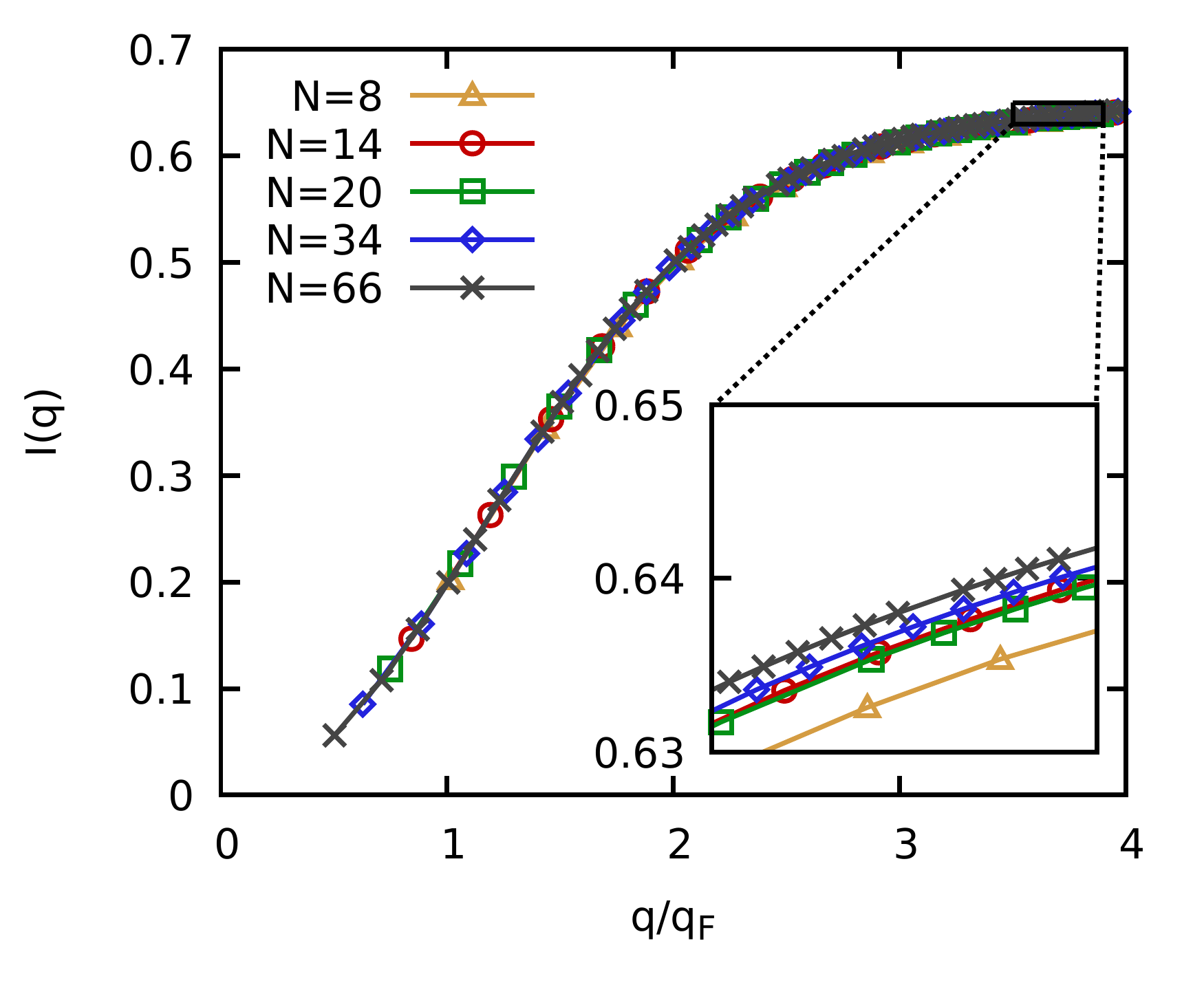}
\caption{\label{fig:Ginfty_rs10_theta1}
PIMC data for the $q$-dependence of the interaction function $I(q)$ [cf.~Eq.~(\ref{eq:define_F})] for $r_s=10$ and $\theta=1$. The different symbols correspond to $N=8,14,20,34,$ and $66$ electrons and the inset shows a magnified segment for large $q$.
}
\end{figure}

Finally, we show results for the interaction integral $I(q)$ [cf.~Eq.~(\ref{eq:I})] in Fig.~\ref{fig:Ginfty_rs10_theta1}. In contrast to $F(q,\tau)$, here there do appear some differences between $N=8$ (yellow triangles) and the other curves. We thus conclude that there should be no finite-size effects in the reconstructed dynamic structure factors except possibly for $N=8$.


\begin{figure*}\centering\includegraphics[width=0.48\textwidth]{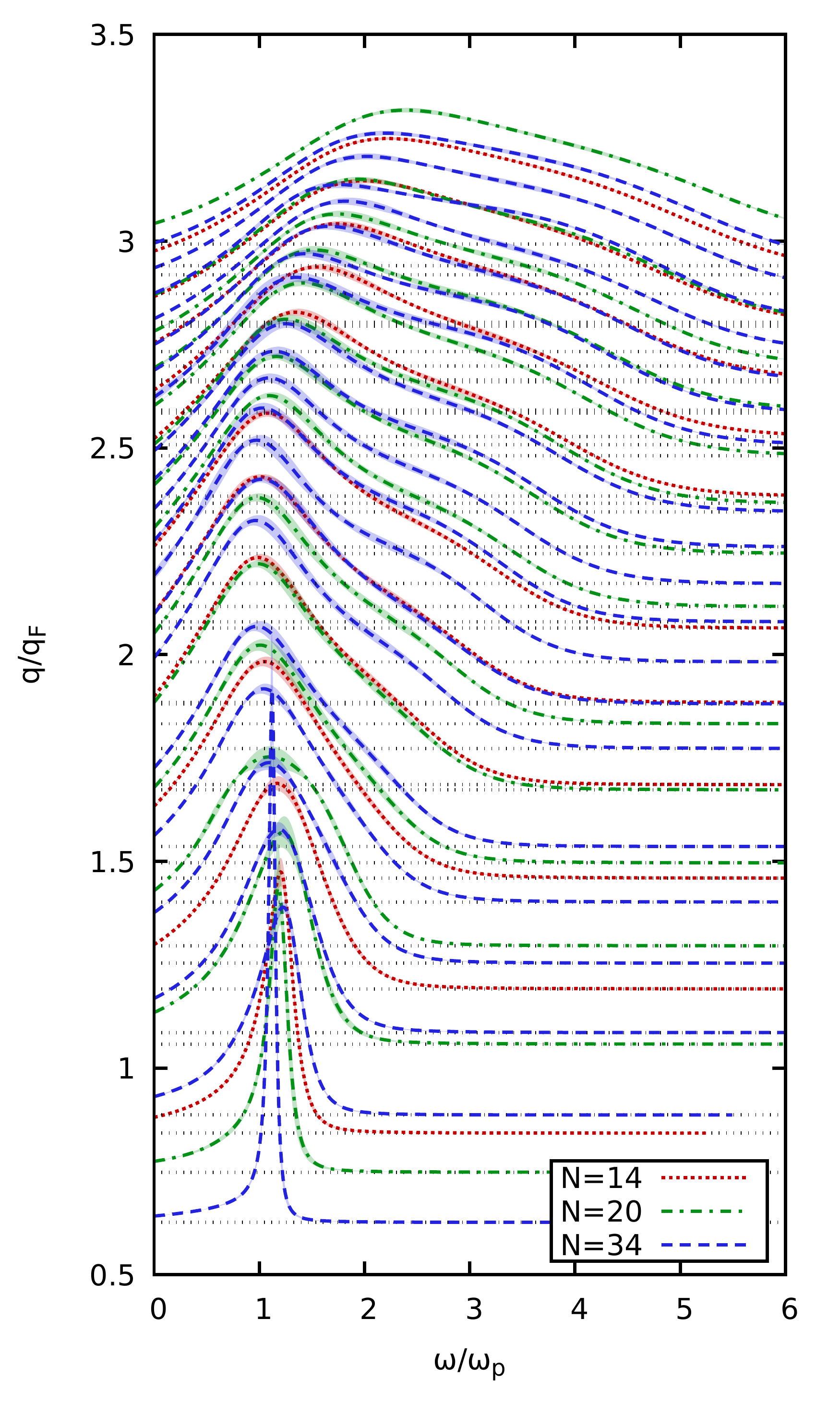}\includegraphics[width=0.48\textwidth]{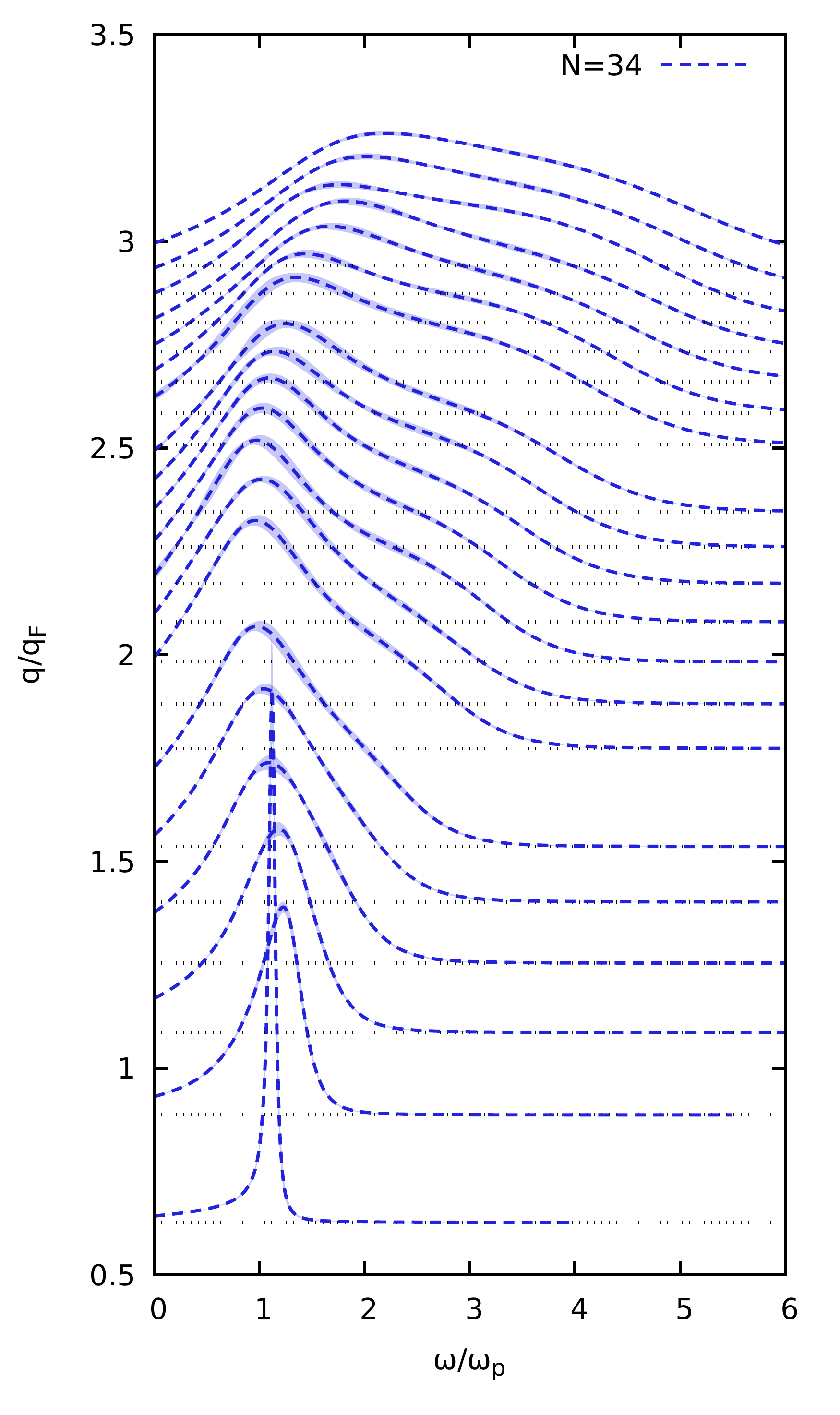}
\caption{\label{fig:Spectral_DSF_rs10_theta1_COMPLETE}
Fully reconstructed dynamic structure factor of the UEG at $r_s=10$ and $\theta=1$ for selected wave numbers. The dashed red, dash-dotted green, dashed blue, and solid black curves have been obtained for $N=14$, $N=20$, and $N=34$ electrons, respectively.
}
\end{figure*}

Let us conclude this study with an analysis of $S(q,\omega)$ itself, which is shown in the $q$-$\omega$-plane for selected wave numbers in the left panel of Fig.~\ref{fig:Spectral_DSF_rs10_theta1_COMPLETE} for $N=14$ (dotted red), $N=20$ (dash-dotted green), and $N=34$ (dashed blue) electrons. Due to the reduced density and the lower temperature compared to the WDM dispersion shown in Fig.~\ref{fig:Spectral_DSF_rs2_theta2_COMPLETE}, the curve for the smallest $q$-value for $N=34$ exhibits a rather sharp peak that is only slightly shifted away from the plasma frequency. In this context, we remark that our reconstruction scheme has no problem with such distinct features, which is in stark contrast to other inversion methods where the obtained spectra are often artificially broadened.
Furthermore, we have obtained the familiar dispersion with the superposition of a mean-field contribution around $\omega_\textnormal{mf}=q^2/2+\omega_\textnormal{p}$ and an additional incipient mode at lower frequencies that has been reported in Ref.~\cite{dornheim_dynamic}.

Yet, the important point for the present investigation is that no systematic finite-size effects occur between the curves for different $N$ at subsequent wave numbers. For completeness, we note that there do occur some small variation for some larger wave numbers, these are an artifact of the reconstruction procedure itself and not related to $N$. To verify this claim, we also show the full dispersion relation (i.e., all $q$-values in the depicted wave-number range) in the right panel of Fig.~\ref{fig:Spectral_DSF_rs10_theta1_COMPLETE}. Here, one can clearly see that the dynamic structure factor for the third-largest $q$-value is somewhat inconsistent to the other curves, although the system size is the same everywhere.

\begin{figure}\centering\includegraphics[width=0.475\textwidth]{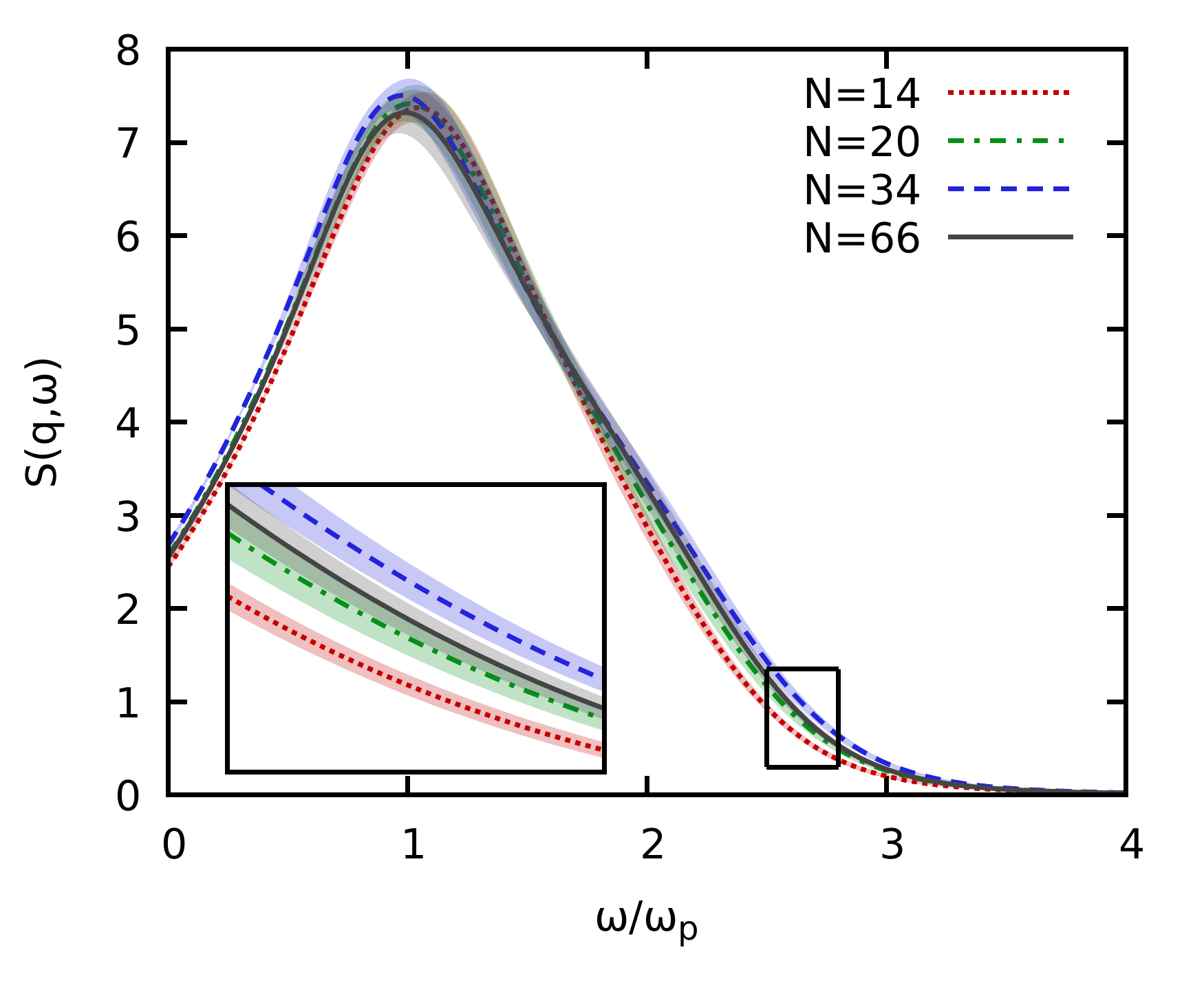}
\includegraphics[width=0.475\textwidth]{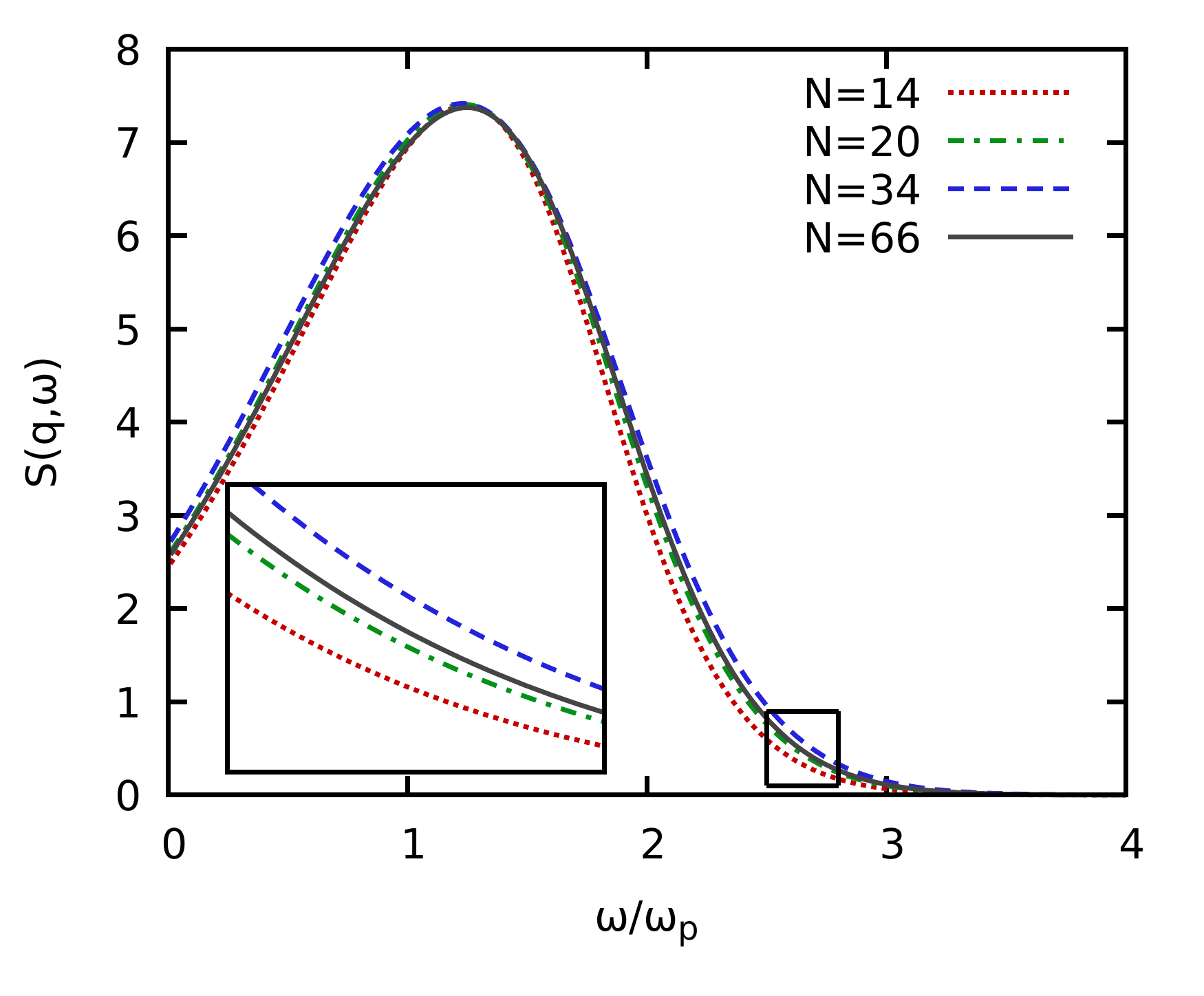}
\caption{\label{fig:DSF_rs10_theta1_multi_N}
Dynamic structure factor of the UEG at $r_s=10$ and $\theta=1$. The dotted red, dash-dotted green, dashed blue and solid black lines correspond to $N=14$ ($q/q_\textnormal{F}\approx1.460$), $N=20$ ($q/q_\textnormal{F}\approx1.496$), $N=34$ ($q/q_\textnormal{F}\approx1.536$), and $N=66$ ($q/q_\textnormal{F}\approx1.508$) electrons, where the numbers in brackets indicate the respective wave number $q$. The left panel shows the full reconstructed solutions for $S(q,\omega)$, and the right panel the corresponding curves from the static approximation that has been included as a reference.
}
\end{figure}

The lack of finite-size effects in the dynamic structure factor can also be seen even more clearly in Fig.~\ref{fig:DSF_rs10_theta1_multi_N}, where $S(q,\omega)$ is shown for different particle numbers at similar wave numbers around $q=1.5q_\textnormal{F}$. The left panel shows the reconstructed solutions [i.e., using the full frequency dependence of $G(q,\omega)$] and, similar to the WDM example depicted in Fig.~\ref{fig:DSF_rs2_theta2_multi_N}, only minor differences occur, mainly for large frequencies.
Again, these small deviations are fully explained by the slightly different $q$-values for the four curves, and completely reproduced by the static approximation shown in the right panel, see also Sec.~\ref{sec:WDM} for a more detailed discussion.

Finally, we mention that here, too, no consistent solutions could be found for $N=8$, which further substantiates our previous finding from the WDM regime that finite-size effects manifest not in an $N$-dependence of $S(q,\omega)$ itself, but in the impossibility to match the exact constraints on $G(q,\omega)$ [cf.~Sec.~\ref{sec:reconstruction}] with the PIMC data.

\section{Summary and Outlook\label{sec:summary}}

In this work, we have investigated in detail the possibility of finite-size effects in the dynamic structure factor of the uniform electron gas both at WDM conditions, and at the margins of the strongly coupled electron liquid regime. More specifically, $S(q,\omega)$ can be accurately reconstructed on the basis of \textit{ab initio} PIMC data, which, while being exact with respect to exchange--correlation effects, have been obtained for a finite simulation cell.
In a nut shell, we have found that even as few $N=14$ electrons are sufficient to give accurate results for $S(q,\omega)$ that are converged with respect to $N$, and no system-size dependence could be resolved within the given confidence level. In contrast, no solutions for $S(q,\omega)$ could be found for $N=8$ electrons, as the exact constraints on the dynamic local field correction $G(q,\omega)$ that are incorporated into the reconstruction procedure cannot be matched to the PIMC data when the latter are not converged with respect to $N$.

Therefore, the current analysis further corroborates the high quality of the electronic structure factors presented in Refs.~\cite{dornheim_dynamic,dynamic_folgepaper}. This is an important finding, as the dynamic density response is of key relevance for many applications (see Sec.~\ref{sec:introduction}) like the construction of dynamic exchange--correlation kernels for time-dependent DFT simulations~\cite{Baczewski_PRL_2016}
or the ongoing investigation of the incipient excitonic mode in the UEG~\cite{dornheim_dynamic}.

\section*{Acknowledgments}

   We gratefully acknowledge computing time on a Bull Cluster at the Center for Information Services and High Performance Computing (ZIH) at Technische Universit\"at Dresden.
   
  This work was partially funded by the Center for Advanced Systems Understanding (CASUS) which is financed by Germany’s Federal Ministry of Education and Research (BMBF) and by the Saxon Ministry for Science, Culture and Tourism (SMWK) with tax funds on the basis of the budget approved by the Saxon State Parliament.

\section*{References}

\bibliography{bibliography}

\end{document}